\documentclass{article}

\newcommand{\vir}[1]{``#1"}

\newcommand{\ignore}[1]{}
\usepackage{arxiv}
\usepackage[utf8]{inputenc} 
\usepackage[T1]{fontenc}    
\usepackage{url}            
\usepackage{booktabs}       
\usepackage{amsfonts}       
\usepackage{nicefrac}       
\usepackage{microtype}      
\usepackage{lipsum}		
\usepackage{graphicx}
\usepackage{doi}

\usepackage{graphics,amssymb,amsmath,epsfig,color}
\usepackage{graphicx}

\title{Digital Transformation in the Public Administrations: a Guided Tour For Computer Scientists}

\author{ 
\href{https://orcid.org/0000-0002-7958-9924}{\includegraphics[scale=0.06]{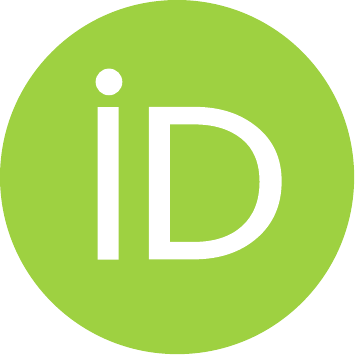}\hspace{1mm}Paolo ~Ciancarini}\thanks{Corresponding author, \href{email:email-id.com} paolo.ciancarini@unibo.it} \\
	Department of Computer Science - Science and Engineering\\
	University of Bologna\\
	Bologna, 40126, Italy \\
	\texttt{paolo.ciancarini@unibo.it} \\
        \And
\href{https://orcid.org/0000-0002-6286-8871}{\includegraphics[scale=0.06]{orcid.pdf}\hspace{1mm}Raffaele ~Giancarlo}\\
	Department of Mathematics and Computer Science\\
	University of Palermo\\
	Palermo, 90121, Italy \\
	\texttt{raffaele.giancarlo@unipa.it} \\
	\And
 \href{https://orcid.org/0000-0001-6001-5050}{\includegraphics[scale=0.06]{orcid.pdf}\hspace{1mm}Gennaro~Grimaudo} \\
	Department of Engineering\\ University of Palermo and Sicily Regional Government \\
	University of Palermo\\
	Palermo, 90121, Italy \\
	\texttt{gennaro.grimaudo@unipa.it} \\
}

\renewcommand{\shorttitle}{\textit{Digital Transformation in the Public Administrations: a Guided Tour For Computer Scientists} }

\hypersetup{
pdftitle={A template for the arxiv style},
pdfauthor={Paolo~Ciancarini, Raffaele~Giancarlo, Gennaro~Grimaudo},
pdfkeywords={digital transformation, public sector, open data, open government data, data governance, privacy and security, smart cities, cloud computing, digital twins, digital shadows, digital skills, leadership, co-creation, agile processes},
}

\graphicspath{./images/}

\begin{document}
\maketitle

\begin{abstract}
        {\textbf{Motivation:} Digital Transformation (DT) is the process of integrating digital technologies and solutions into the activities of an organization, whether public or private. This paper focuses on the DT of public sector organizations, where the targets of innovative digital solutions are either the citizens or the administrative bodies or both. This paper is a guided tour for Computer Scientists, as the digital transformation of the public sector involves more than just the use of technology.  While technological innovation is a crucial component of any digital transformation, it is not sufficient on its own. Instead, digital transformation requires a cultural, organizational, and technological shift in the way public sector organizations operate and relate to their users, creating the capabilities within the organization to take full advantage of any opportunity in the fastest, best, and most innovative manner in the ways they operate and relate to the citizens. }    \\
        
        {\textbf{Results:} Our tutorial is based on the results of a survey that we performed as an analysis of scientific literature available in some digital libraries well known to Computer Scientists. Such tutorial let us to identify four key pillars that sustain a successful DT: (open) data, ICT technologies, digital skills of citizens and public administrators, and agile processes for developing new digital services and products. The tutorial discusses the interaction of these pillars and highlights the importance of data as the first and foremost pillar of any DT. We have developed a conceptual map in the form of a graph model to show some basic relationships among these pillars. We discuss the relationships among the four pillars aiming at avoiding the potential negative bias that may arise from a rendering of DT restricted to technology only. We also provide illustrative examples and highlight relevant trends emerging from the current state of the art.}    \\
\end{abstract}


\section{Introduction}\label{sec:intro}

We all are citizens of a digital era, which offers new possibilities, new rights, new duties \cite{isin2020being}.  Digital citizens use ICT technologies to communicate, access information, and participate in social, economic, and political activities. The impact that the  advent of this new era has had is not limited to citizens, but also involves public administrations (\emph{PAs}, for short),  being directed to be more and more digital in how their work, relate, and interact with citizens.
As a matter of fact, a Digital Transformation (\emph{DT}, for short) in the \emph{PAs} is taking place. To date, there is no formally accepted definition of such a transformation: as pointed out in \cite{mahraz2019systematic}, there can actually be many of them. In the context of this work, we use the definition below that attempts to summarise most of them. \emph{DT} is the process of integrating digital technologies and solutions into all aspects of the activities of an organization, whether public or private. For this reason, it can be a complex, never-ending, and often discouraging process.

DT represents a cultural, organizational, social, and technological shift that leads organizations to initiate a change in the way they operate and relate to their users, trying to be more responsive to their needs. It must have an associated strategy, focusing on creating the capabilities within the organization to take full advantage of the opportunities of new technologies and their impact in the most useful and most innovative manner, keeping in  mind that the private and public sector differ. Indeed, the technologies used in the private sector cannot be applied immediately in the public one  without an analysis of possible differences of impact \cite{JLA22}. This is due to the fact that  \emph{DT} in the private sector concerns how it impacts employees and how digital technologies and processes improve the productivity and quality of the products offered to customers. In the public sector, \emph{DT} has a different scope because it impacts not only the citizens, who are scarcely comparable to customers, but also the governing and administrative processes, and the nature itself of the social contract \cite{vial2019understanding}. From now on, unless otherwise specified, in our paper \emph{DT} refers specifically to the public sector.

As digital transformation projects are implemented, citizens become eager for accessing digital services supporting their activities and life. This requires public institutions to ensure that their digital solutions are user-friendly, secure, and accessible to all citizens \cite{eucomm2016}. Several institutions and administrations in the public sector are  exploring the opportunities offered by  digital transformation technologies to enhance their organizational flexibility necessary to adapt to changing contexts and meet new government and citizens demands \cite{danielsen2021benefits, leao2018digitization}. More in general, \emph{DT} has become an increasingly pressing issue in recent years, with the growing need to modernize government services and improve their efficiency and transparency with respect to the needs of citizens. 
It is not surprising that there are several areas in this context where Computer Science can have a major impact, while receiving further stimuli for its growth. One of them  is the development and implementation of digital infrastructures to support specifically the public sector, with particular attention to privacy and security aspects of the data and of their processing. 
This includes creating systems for data collection, storage, and analysis \cite{Otto2022}, as well as building networks and platforms for communication and collaboration among different public institutions \cite{vaira2022smart}. Another one is the study regarding how to make government data more accessible, transparent, and useful for the citizens, their administrations, and other stakeholders. This includes developing standards for data sharing and interoperability, as well as creating tools and applications that enable citizens to engage with public Open Data in useful ways \cite{nikiforova2021open}.

Therefore, \emph{DT} is certainly of interest for Computer Scientists but, to the best of our knowledge of the State of the Art, the complexity of the \emph{DT} and the interplay among its key aspects does not seem to be well presented to a Computer Science audience. Filling this gap in the Literature is the aim of this tutorial.  Indeed, based on the State of the Art, we identify four key pillars that sustain a successful \emph{DT}. Specifically, (open) data, ICT technologies, digital skills of citizens and public administrators, and agile processes. We dedicate part of this tutorial  to the description of each and another part to their interaction. Indeed, although ICT technologies are essential in driving any digital transformation, and well known to Computer Scientists, they are not enough on their own. Therefore, our  aim is to present the benefits of technology and avoid the potential negative bias that may arise from a rendering of \emph{DT} restricted to technology only. This is a novelty for a presentation of this area.

Another novelty is that we focus on data as the first and foremost pillar of any digital transformation. The case of the public sector involves the integration of data technology in all aspects of governance to improve efficiency, transparency, and citizen engagement. Open Data is a critical component of this transformation as it enables the public sector to make its data publicly available for reuse and re-purposing by others. In particular, Open Data in the \emph{PA}  refers to the idea of making government data available to the public in a usable and accessible format. 
 They  can include information on government spending, public transportation, healthcare, education, environmental issues, and much more. Their availability has the aim to allow  citizens to better understand how the government operates, and how it spends public money. This increased transparency can improve public trust, accountability, and collaboration between the government and its citizens. It also provides valuable insights to policymakers, researchers, and other stakeholders. The corresponding implementation of Open Data policies is crucial for unlocking the potential of data in the \emph{PA}. 

The use of ICT technologies, including smart cities and data governance of public clouds, can help to improve the delivery of public services \cite{yukhno2022digital}. 
While smart cities leverage technology to improve quality of life, reduce costs and improve the sustainability of urban areas, the data governance of public clouds can provide a cost-effective and scalable infrastructure for public administrations, allowing them to deliver and monitor their services more efficiently and effectively. We remark that citizens must have the necessary competencies to access and use digital services \cite{liu2022citizen}. 
This encompasses digital literacy, data literacy and online security awareness. Governments need to invest in programs that promote digital skills development, particularly for underprivileged and marginalized groups. Digital transformation is an ongoing process of integrating technology into various aspects of society, and digital  competencies of the citizens play a crucial role in driving this integration. Citizens must be able to navigate the risks associated with online activity and understand the importance of protecting their personal information.

Beyond basic digital skills, citizens should also have a strong understanding of the role of technology in the public sector, and the benefits it can bring to government services. This requires a strong understanding of public policy, as well as an awareness of the needs and interests of different stakeholders. An active digital citizenship also requires a willingness to collaborate and work with others. This includes engaging in online communities, collaborating on digital projects, and building partnerships with government and other stakeholders to co-create digital contents and services.

Finally, we show why we believe that any \emph{DT} is a process that should exploit an agile approach, when introducing or adopting new services for citizens. Such an approach allows for rapid development and testing of new services, with continuous feedback and improvement. Since \emph{DT} is an iterative process that requires continuous adaptation  to the needs of the citizens and improvements in the policies  of the administrators, Agile processes are the ideal \vir{operational tool} for \emph{DT}. 

The scope of this tutorial includes public administrations such as municipalities, national governments, and other governmental bodies. We exclude other specific public sector institutions such as the military and homeland defense organizations, educational organizations like universities, or health organizations like hospitals. Our primary goal is to explore how these public administrations can leverage digital technologies to improve their services, operations, and interactions with their citizens. 
 
This tutorial has the following structure:
Section \ref{sec:methodology} describes how we selected the literature at the basis of this tutorial (after checking that no similar tutorial has been published);
Section \ref{sec:guidedtour} introduces a graphic map of the tutorial, which helps to clarify it structure.
Section \ref{sec:Res_Data} describes the data ecosystem that is at the basis of the digital information of the public sector.
Section \ref{sec:Res_Technology} describes two technological subareas especially relevant, namely Smart Cities and (open) Data Governance.
Section \ref{sec:Res_People} discusses  the technical aspect we have found concerning People that got our attention, based on the Literature search: digital skills and citizens' co-creation of contents and services.
Section \ref{sec:Res_Process} presents  the two main technical aspects we have found relevant of the \emph{DT} processes: Change Management and Frameworks and Maturity Models.
Section \ref{sec:Future_Directions} contains the main discussion of the major ideas presented in this tutorial.
Finally, the last Section \ref{sec:Conclusions} is a wrap-up of this tutorial.

\section{Literature Selection}\label{sec:methodology}

    Our effort to provide a systematic homogeneous presentation of \emph{DT} is based on the current State of the Art. Therefore, a first essential step is to resort to established methods to collect relevant papers for a Literature Review. However, while the former describes relevant papers covering the State of the Art, this Tutorial uses the selected papers to extract the main ingredients of \emph{DT} and to propose models for it, with the addition of illustrative examples. Details regarding the paper selection process follow. 
    
    We have performed a literature search in the ACM and IEEE digital libraries, in addition to Google Scholar. Given that, as stated in the Introduction, the focus is on Agile methodologies, the query term is very focused: \vir{agile} AND \vir{digital transformation} AND \vir{public services}. The period of time is January 2017-February 2022.  
        
    The search outcome is  as follows: 1,992 references from  Google Scholar, 745 from IEEE, and 3,143 from ACM Digital Library. For ACM, only the first 2000 items (sorted by relevance) could be accessed, since the search engine limits itself to report that the bottom 1,143 ones are very similar to those available for display. Therefore, over the three databases we have consulted, we have collected a total of 5,880 papers.
    
    After a more detailed review of the titles and abstracts, we selected 151 papers for our initial \vir{core collection}, as they primarily focused on computer science technical content related to DT, as opposed to social and ethical issues.  
    During the reading phase of the core papers, it became  evident that some additional literature was required, in order to make the Tutorial more comprehensive. This included 13 references covering technical background topics (such as data organization in \cite{Otto2022}), 36 references covering reference standards, models, and regulatory procedures specific to DT (such as DT maturity models in \cite{CMMI} and data protection regulation in \cite{GDPR}), and 41 more recent papers that were not included in our initial core collection (so extending the core to 192 papers). In total, we collected an additional 90 papers.

\section{The Main  Ingredients of the Digital Transformation in the Public Sector} \label{sec:guidedtour}
 
   Given the definition of \emph{DT} provided in the Introduction, we present here its main ingredients  in terms of four domains that emerge from an analytic reading of the papers we have considered for this research. We also discuss the interactions among them. The end result is a graph model of \emph{DT}, proposed here for the first time, and that can be used as a \vir{summary map} to describe the \emph{DT} process. Moreover, the concepts and notions summarized by the model  are exemplified via  two paradigmatic examples: the Cities of Barcelona and Chicago. Such a choice is motivated by the fact that, although complex cities, their  \emph{DT} scale  is well suited for the crisp identification   and evaluation of the  specific actions regarding their transition to digital.

    \subsection{A Graph Model for DT} \label{sec:Panoramic_Building_Blocks}

    From an examination of the 192 core papers 
    although they address various aspects related to how \emph{PAs} plan and implement their \emph{DT} strategies, they predominantly focus on one of four knowledge domains: {\bf{\emph{Data}}} (36 papers), {\bf{\emph{Technology}}} (61 papers), {\bf{\emph{People}}} (33 papers), and {\bf{\emph{Process}}} (62 papers). These domains are briefly discussed below.
    
    \begin{itemize}
    
        \item{ {\bf{\emph{Data}}}. The availability of public data has changed significantly over the past decade, resulting in a greater awareness of how it is collected, represented, owned, and managed. As a consequence, the data life-cycle has changed with respect to the past \cite{attard2015systematic, filograna2016cloudification, i2017government, margariti2020assessment, saba2022toward}, posing new technical problems even to mature areas such as databases \cite{kraska2022seattle} and requiring new ways to design software for their management \cite{davoudian2020big}. As far as this Tutorial is concerned, it is important to point out that data are no longer seen as an asset to exploit for a competitive advantage, but as a social \vir{infrastructure} that must be made available to policy makers and citizens  to ensure and improve the well-being of Society \cite{harrison2019applying, hastings2019unlocking}. With this in mind, more and more \emph{PAs} are making available their data to improve transparency and accountability \cite{bounabatgovernment, i2017government, ortiz2021design, rong2022smart, toots2017framework, ylipulli2020smart}. However, due to the heterogeneity and  lack of interoperability of the data sources, major problems arise. One is how to exploit at its best the information contained in those data. Another is the realization of the sound technical principle of \vir{only once}, i.e., data collected by one administration  should be available to other administrations. Scale factors make these problems even more difficult, since \vir{data} can refer to a continent \cite{kalvet2018contributing}, a nation, a city, or be sector specific \cite{malhotra2019designing, kalogirou2020linked}. In order to address the problems alluded to earlier and of which we have provided two examples, an entire data ecosystem is shaping up, ranging from infrastructures to data analysis tools and applications. Following \cite{Otto2022}, the term \textit{ecosystem} is used here, instead of environment, because like real ecosystems, data ecosystems are designed in such a way to have an \vir{evolutionary} part aimed at improving data quality levels over time.  It is a node of the proposed graph model and, in what follows, we use the terms data ecosystem and data interchangeably. Moreover, being data the source of information  that powers the \emph{DT}, its corresponding node is the central one in the model, as shown in Figure \ref{fig:DT_graph_model_main}. Additional details regarding the components of such a node are presented in Section \ref{sec:Res_Data}.}
        
        \item{ {\bf{\emph{Technology}}}. The term technology refers to hardware and software systems supporting \emph{PAs} in some \emph{DT} process \cite{battisti2020digital, bounabatgovernment, cordella2019government, panagiotopoulos2019public}. Digital platforms that support all stages of governance activities are in place or planned  \cite{vaira2022smart}, since their realization is perceived as a way to increase the pace of the \emph{DT} \cite{mydyti2020cloud}.  In particular, several \emph{PAs} are moving to the Cloud \cite{abied2022adoption, chaudhuryreforming, filograna2016cloudification, pinheiro2020towards}. Smart cities \cite{saba2022toward} are becoming a recurring pillar in the \emph{DT}. Blockchain technologies are also being considered, but they appear somewhat marginal at this stage \cite{wingren2021blockchain}. Artificial intelligence is expected to play a major role, e.g, \cite{jadi2017implementation, neumann2022exploring, saura2022handbook, tamburri2020dataops, van2020evaluating, viscusi2020governments}, although its impact and pervasiveness on privacy, transparency and accountability in the realm of \emph{DT} is still under study \cite{vaira2022smart}. Such technologies are useful for supporting data-driven decision-making in public  administrations which, in turn, have the goal to provide  a higher and higher  quality of life for their citizens. In order to achieve this  goal, in particular for limited geographic areas such as cities, ruling bodies and decision makers  are accepting difficult and stimulating challenges related to the creation of complex digital models of cities, that would allow them to respond to the needs of citizens faster than in the past (e.g. \cite{saba2022toward}). Those new models must ensure privacy and security, making necessary for the \emph{PAs} to possess regulations about data governance, e.g., the European Data Protection Act (\emph{GDPR}) \cite{GDPR, datta2020digital}, cyber-security technologies, e.g., the National Cyber-Security Agencies \cite{Nat_CS_Agency_IT, Nat_CS_Agency_ES, Nat_CS_Agency_US}} and a flexible and modular strategy to data access and sharing, e.g., the European DECODE project \cite{DECODEproject}. Technology for the \emph{DT} is the node shown in Figure \ref{fig:DT_graph_model_main}. Details regarding the components of such a node are presented in Section \ref{sec:Res_Technology}.

        \item{ {\bf{\emph{People}}}. The services provided by the \emph{PAs} must be considered valuable by the citizens, who sustain them by paying taxes.
        Such a fact has an important consequence regarding efficiency, which has had a privileged position in the design and deployment of services: the aim for it, although valuable, is by no means sufficient in generating services perceived of value to the citizens \cite{cordella2019government}. In fact, for the design of services, a people-driven delivery model is more and more the one of choice \cite{battisti2020digital, bounabatgovernment}. Such a new model places citizen participation at the center of most service design and implementation initiatives, whose success must be evaluated by their users, namely the citizens and various kinds of decision-makers, according to their perception of the value created \cite{agbozo2020towards, mcbride2019does, mergel2018citizen, nachit2021digital, scrupola2021value}. Interestingly, 
        although the meaning of services valuable to the citizens is clear, the meaning of the  apparently related term of \vir{business value} in the digital \emph{PA} is not so clear, although intuitively it relates to the provision of better service to the citizen, efficient operation of the services and the enforcement of  the law  \cite{lopes2018business}. People for the \emph{DT} is the node shown in Figure \ref{fig:DT_graph_model_main}. Details regarding the components of such a node are presented in Section \ref{sec:Res_People}}.
        
        \item{ {\bf{\emph{Process}}}. As part of a \emph{DT} strategy whose aim, as already stated, is to  obtain services that are more citizen-centered, it is to be expected that new ways of process engineering are developed and deployed \cite{aleinikova2020project}. Although the transition from established process engineering to new ones is not so simple \cite{masombuka2020framework, wipulanusat2019drivers}, Agile project management approaches seem to be the \vir{best} candidates to support the mentioned transition \cite{aleinikova2020project, bogdanova2020agile, bounabatgovernment, chaudhuryreforming, jonathan2019digital, mantovani2020characteristics, mubarkoot2021assessment, nachit2021digital}.  Another crucial aspect regarding process management is  the need for  new ways  to measure success, i.e., in terms of a meaning of \vir{value} that is certainly application specific, but with a rather broad spectrum. Rather than being specific on those measurements, the trend is to measure the degree of maturity achieved by the processes in the \emph{DT} strategy implementation. In this regard, in the Literature available in this Tutorial, we find several papers proposing different Frameworks and Maturity Indexes, e.g., \cite{bakar2020digital}, with which stakeholders could measure the progress achieved in the digital transition of their Organizations. However, among the many available, the GovTech Maturity Index (GTMI, for short)  proposed by the World Bank \cite{dener2021govtech} seems to be the most reliable one.      
        Process for the \emph{DT} is the node shown in Figure \ref{fig:DT_graph_model_main}. Details regarding the components of such a node are presented in Section \ref{sec:Res_Process}}.

    \end{itemize}

    \subsection{Interactions Among Knowledge Domains}

\begin{figure*}[!ht]
        \centering
		\includegraphics[scale=0.5]{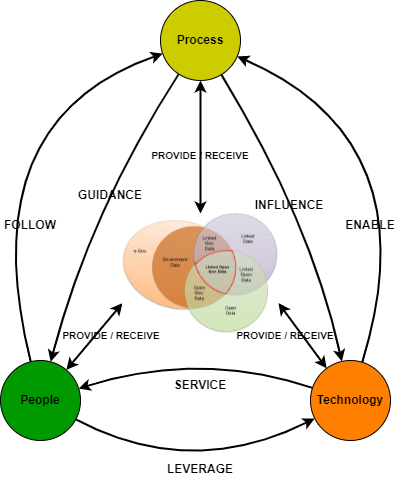}
		\caption{{\bf The \emph{DT} Graph Model}. Nodes represent the knowledge domains.    Each edges represents interaction and dependencies between its end nodes, while the label on each edge indicates the type of relationship between its nodes, following the main text.}
		\label{fig:DT_graph_model_main}
    \end{figure*}

    The papers we reviewed show that the four knowledge domains presented above have several mutual interactions and dependencies, summarized in terms of edges in the graph model in  Fig. \ref{fig:DT_graph_model_main}. Each edge between two nodes (knowledge domains) encodes an interaction between its end-points, while the direction of the edge encodes the dependence, i.e., an edge $(a,b)$ indicated that $a$ depends on $b$ with the label  indicating the nature of such a dependency. Details are provided next.  
    
    \begin{itemize}
    
        \item {\bf{Interactions with \emph{Data}}}. At the heart of our graph model there is the data ecosystem,  subject to incoming and outgoing data flows. In one direction, \textit{data} is the source of information for other nodes, and in the other direction, it grows from the information it receives from other nodes. In the graph model, we encode this bi-directionality in terms of a  \emph{provide/receive} paradigm. People provide data and receive information and know-how. Processes provide business intelligence, statistics, etc, and receive data. Technologies provide tools for a better governance over the data, and receive data. It should be emphasized that this paradigm encodes well the flow of information in a  data ecosystem \cite{attard2015systematic}.

        \item {\bf{Interactions with \emph{Technology}}}.
        Technologies are at the service of the citizens. Indeed, given that a needs-based holism means the reunification of government services around citizens rather than business processes \cite{margetts2013second}, these technologies enable it, with the end results that increase the capacity of \emph{PAs} to respond to the emerging needs of citizens \cite{al2022healthcare, vestues2021usingProcess}.
        Technologies also enable processes and facilitate their re-engineering phase \cite{ylinen2021digital}.

        \item {\bf{Interactions with \emph{People}}}. People, who follow the processes and initiatives provided by \emph{PAs}, leverage different emergent technologies \cite{ylinen2021digital} to enable the improvement of public processes (\emph{process re-engineering}).

        \item {\bf{Interactions with \emph{Process}}}. Processes guide people, through the provision of quality digital services, to respond more effectively and promptly to the evolving needs of citizens \cite{syahrizal2020finding}. At the same time, technological choices are often influenced by the processes, and how these are designed or re-engineered \cite{ylinen2021digital}.
        
    \end{itemize}
    
    \subsection{Accounting for  Responsiveness} \label{sec:Tour_Responsiveness}

\begin{figure*}[!ht]
        \centering
		\includegraphics[scale=0.30]{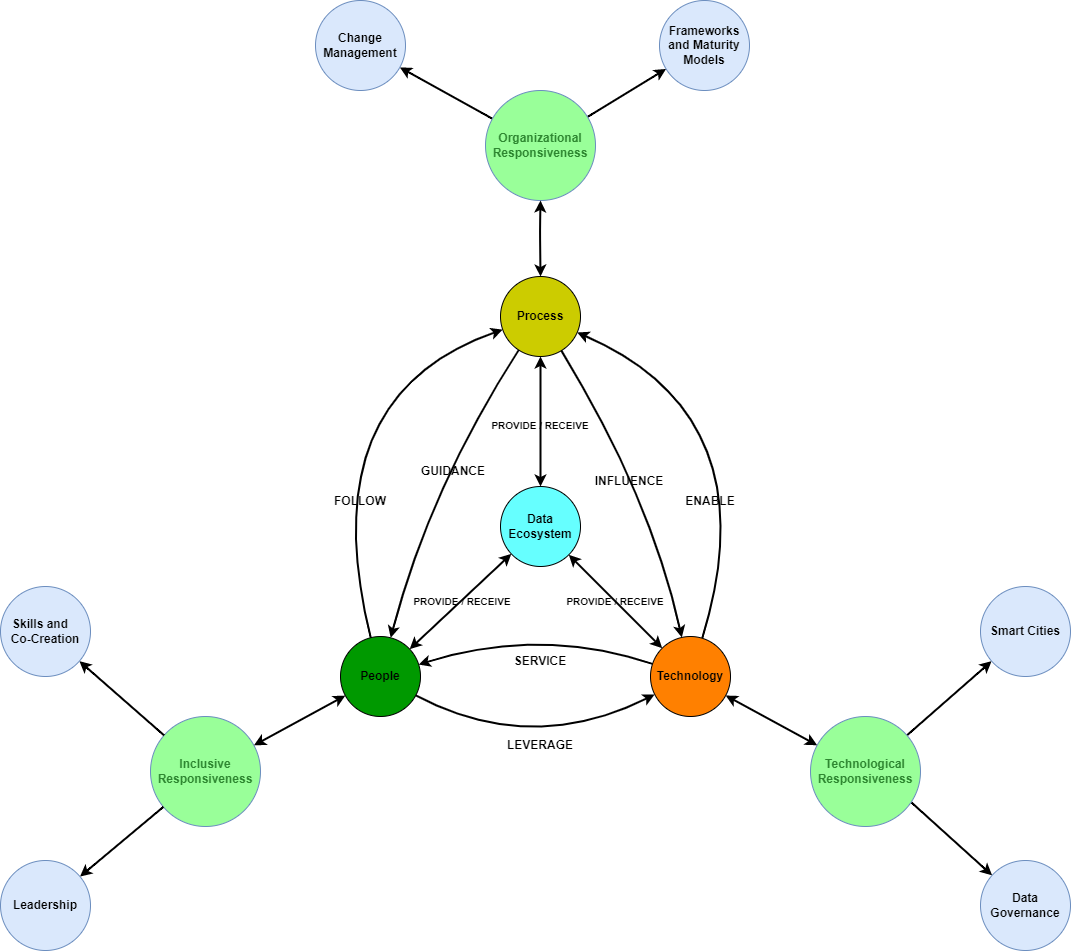}
		\caption{{\bf \emph{DT} Graph Model Augmented with Responsiveness Terms}. The \emph{DT} Graph Model in Figure \ref{fig:DT_graph_model_main} is augmented in correspondence of the Knowledge Domains. Namely, {\bf{\emph{Technology}}}, {\bf{\emph{People}}} and {\bf{\emph{Process}}}. {\bf{\emph{Technology}}} is augmented with the Technological Responsiveness aspects, such as Smart Cities and Data Governance. {\bf{\emph{People}}} is augmented with Inclusive Responsiveness aspects, such as Skills, Co-Creation, and Leadership. {\bf{\emph{Process}}} is augmented with the Organizational Responsiveness aspects, such as Change Management and Frameworks and Maturity Models.}
		\label{fig:DT_graph_model}
    \end{figure*}

    As stated in the Introduction, one of the main goals of \emph{DT} is to increase the responsiveness  of an organization to the  changing needs of citizens. It is evident that in response to changes, such a goal can be reached by being able to: (a) quickly use novel technologies; (b) implement   an  inclusive strategy that promptly makes available 
    new skills to citizens, administrators  and policy-makers; (c) adopt flexible  organizational models for the  design, implementation  and deployment of  services. 
    Those  main aspects of responsiveness in \emph{DT} can be summarized as follows: technological responsiveness, which naturally connects to the  knowledge domain of {\bf{\emph{Technology}}}; inclusive responsiveness, which naturally connects to the knowledge domain of {\bf{\emph{People}}}; and organizational responsiveness, which naturally connects to the knowledge domain of {\bf{\emph{Process}}}. Therefore, it is felt appropriate to extend the graph model  proposed here accordingly, as in Figure \ref{fig:DT_graph_model}. We now discuss the terms we have just introduced.

    \begin{itemize}
    
        \item{\bf{Technological Responsiveness}}. It concerns the flexibility and versatility of the solutions adopted for the collection, representation, and management of the data, together with the appropriate infrastructures to host and manage them \cite{alghamdi2017desinge, rakhman2019design, tangi2020barriers, vestues2021usingPeople, vestues2021usingProcess}. Those solutions must account for good levels of quality and privacy. The meaning of Quality is given via a set of properties to which data should respond. Specifically: accuracy, completeness, consistency, timeliness, validity, and uniqueness \cite{dai2016data}. As for privacy, in addition to the meaning given to it in the domain of IT security, the solutions granting it must be compliant with current legislation, e.g.,  the European General Data Protection Regulation (GDPR) \cite{GDPR}. A particularly important and novel aspect of data processing is to account for the requirement that  users must be given the option to decide who can process their data and for which purposes. As for computer architectures, to date, there are many of them supporting technological responsiveness in the \emph{DT} \cite{baheer2020systematic}  and even new ones have been proposed, although it is not clear how widespread their adoption is \cite{agarwal2017enterprise}. Moreover, the possibility of migrating from monolithic systems offering services to microservices technologies is also considered \cite{luz2018experience}: although this suggestion is somewhat isolated, the results are encouraging.  From our Literature Review, and in regard to the achievement of responsiveness in the \emph{DT}, it is evident that Smart Cities are technologically very promising and popular, while Data Governance issues are more delicate and difficult for public administrations. Therefore, among the many facets characterizing  technological responsiveness, we concentrate on those two, which are briefly discussed next. 
        
        \begin{itemize}
        
            \item { {\bf{Smart Cities}}. 
            According to the ISO/IEC \cite{ISOIEC_jtc1}, (but see also \cite{moustaka2018systematic}) a Smart City is \vir{an innovative city that uses ICT and other means to improve quality of life, the efficiency of urban operation and services, and competitiveness, while ensuring that it meets the needs of present and future generations with respect to economic, social and environmental aspects}. Moreover, based on a Literature Review, including both academic papers and practical tools, a proposal regarding the key components that make a City smart has been made in \cite{gil2015makes} and validated in \cite{lanza2020what}, specifically for Brazil. The components structure that comes out is hierarchical, with the top level consisting of (a) government; (b) society;  (c) physical environment and (d) technology and data. A second level follows, e.g., point (d) is further detailed into (d.1) ICTs and other technologies, (d.2) data and information. A third level  concludes the hierarchy, e.g., point (d.2) is further broken into (d.2.1) data management, (d.2.2) information processing, (d.2.3) information sharing and integration. 

            Technology is essential for the sustainable development of a smart city (see above and  \cite{rani2021amalgamation}), in particular Internet-of-Things (IoT)  approaches  - see for instance \cite{lepekhin2019systematic, luckner2020urban, javed2020biotope, hsiao2019elevated}. However, technology alone is not enough \cite{calderon2018smartness}. Indeed, starting from the fact that a difficulty for the realization  of a smart city is the fragmented understanding of the interaction between Information Technologies and novel city governance models \cite{pereira2020governance, reis2021exploring, reis2021governance}, changes involving public administration and management seem to be required. For instance,  
            project and risk management  need to be changed: the realization of the infrastructural innovations required to transform a city into a smart one need to be planned carefully in order to avoid delays and over-spending \cite{volpe2022supporting}. Moreover, there is a  need to rethink how software-intensive services are used, in order to implement more flexible infrastructures   \cite{monge2022new, rong2022smart, saba2022toward}. 
            
            Section \ref{sec:Tech_SC_IoT} is devoted to this topic, with a focus on Digital Twins \cite{DWZ2021}, which is a new and promising approach to design and implement a smart city,   based on the a virtual representation of its main physical city objects, including the inhabitants, that interacts with the real objects  and evolves with them \cite{DWZ2021}. For completeness, we point out that Digital Twins are not a new concept, having been introduced by Greives in 2002 and have been the object of rigorous studies in order to identify their range of application domains \cite{barricelli2019survey, fuller2020digital}. }

            \item { {\bf{Data Governance}}. For data governance, it is meant a set of processes, roles, policies, standards, and metrics useful for controlling data management \cite{abell2021cloud, i2017government, monge2022new, rong2022smart, toots2017framework, vijai2020cloud}. Via the effective and efficient management of the amount of structured and unstructured information coming from a multitude of \emph{PA} processes and procedures, its goal is to transform those data into a strategic asset, serving the citizens while preserving their privacy. The issue of data governance is so important and strategic that a new professional figure is emerging:  Chief Data Officer, with its role and responsibilities still being the object of study \cite{nie2018chief, ruslan2022applying}. 
            Certainly, such a figure should be able to manage issues regarding  privacy, security, regulatory compliance, access control, and the resolution of problems caused by poor data quality across the data life-cycle \cite{attard2015systematic, bakar2020digital, datta2020digital,jonathan2020privacy}. Section \ref{sec:Tech_CyberSec} is devoted to Data Governance. }
                   
        \end{itemize}

        \item{\bf{Inclusive Responsiveness}}. It concerns how fast and broad are the cultural changes associated to the acquisition of multidisciplinary skills, ranging from digital to managerial, aimed at gaining greater awareness of the efforts of \emph{DT} \cite{vestues2021usingPeople, vestues2021usingProcess}. Although inclusive responsiveness can be further divided into many categories, here we concentrate on some important ones, i.e., skills development, co-creation, and leadership.  
        We point out that skills development and co-creation are treated synergistically here, inspired by a case study regarding the city of Chicago \cite{mcbride2019does}, which justifies this approach.

        \begin{itemize}

            \item{\bf{Skills and Co-Creation}}. Skills development is a well known concept that needs no further elaboration. Co-creation is a concept that strongly depends on team-building and on the digitization culture that, together with correct communication, enables the actors involved to work together to produce public services successfully \cite{rosler2021value, scrupola2021value, toots2017framework, vestues2021usingProcess}. It is a continuous improvement process, in which \emph{PAs} must implement the necessary tools to successfully exploit feedback from the citizens in the evolution phase of a service \cite{agbozo2020towards, monge2022new, ylinen2021digital}. This approach changes the way in which public services are evaluated, placing the users at the \vir{center}. Indeed,  following earlier research regarding how to measure service quality offered by the \emph{PAs} \cite{arias2018digital}, models and procedures for such a novel \vir{user-centered} evaluation of public services are being investigated \cite{de2019evaluating, menezes2022evaluation}, together with models that identify possible areas, ranging from architectures to risk management, whose improvement would result in the deployment of better services \cite{hermanto2020improving}.  Section \ref{sec:People_Skills_Cocreation} is devoted to this topic.

            \item{\bf{Leadership}}. It is perceived as a fundamental pillar driving \emph{DT} in organizations, including the \emph{PAs} (see \cite{manda2021leadership} and references therein), in particular regarding the definition and implementation of mechanisms that strengthen the governance of digital and smart societies. Although, as pointed out in \cite{kane2015strategy}, strategy rather than technology is the key to success in \emph{DT}, according to the study in \cite{legner2017digitalization}, \emph{PAs} that have reached a certain degree of maturity in the \emph{DT} process are quite likely to have had the support of their managers and their involvement in the formulation of \emph{DT} strategy plans to create new public value. Therefore, IT managers and leaders still play a fundamental part regarding innovation, even with respect to \emph{DT}, but they must also have a deep understanding of which organizational culture is most effective, depending on the type of innovation being implemented \cite{pittenger2022transformational}. In addition, they must have knowledge and training in regard to a specialized set of skills on modern technologies and related cultural changes \cite{tangi2021digital, wipulanusat2019drivers}. Indeed, the current level of expertise, related to emerging technologies, is a barrier to the adoption of these technologies \cite{legner2017digitalization}, while for the creation of services perceived of value critically depends on the level of competencies that managers and decision-makers have regarding technology \cite{luciano2020role}.  Furthermore, managers should behave more like product owners of the new services aiming at meeting the needs of citizens \cite{ghimire2020scaling, kupi2021late, masombuka2020framework, mergel2021agile, mohagheghi2021organizational, monge2022new, ylinen2021digital}.  Yet another key to speed up \emph{DT} is a coordinated policy involving National State, Local States and Municipalities \cite{rodrigues2021impacts}.

            Interestingly, a technological framework based on Digital Twins has been proposed  to help IT governance  \cite{poels2022dt4gitm}. The framework, denoted  Digital Twin for Governed IT Management (\emph{DG4GITM}, for short), links the management of three interconnected systems: IT governance processes, IT management processes, and IT organizational assets by leveraging the technology of Knowledge Graphs and the resulting computational infrastructure. In particular, a given city virtual entity is created through an enterprise ontology "\emph{GITM Domain Ontology}" that is connected to the organization via data flows to populate it with real data from the resources of the organization.

            This point is not the object of further discussion, since we have accounted for all the papers that cover this subject and that we have included in the Literature review. 
    
        \end{itemize}

        \item{{\bf{Organizational Responsiveness}}}. 
        It concerns the ability to adopt rapid organizational changes and to undertake new ways of operating within the \emph{PA} \cite{gong2020towards, maroukian2020leading, mohagheghi2021organizational, tangi2021digital}. \emph{DT} is a continuously evolving process that needs to be monitored in order to evaluate its progress and to identify directions for improvement \cite{maroukian2020leading}. Two key features to consider are  Change Management and Frameworks, and Maturity Models:
        
        \begin{itemize}
        
            \item { {\bf{Change Management}}. It refers to the ability to accept innovation while producing quality services \cite{tangi2021digital, vestues2021usingPeople}. In the \emph{PA} context, one of the essential parts of this point is the promulgation of laws, regulations, and guidelines, which promote the use of the services offered, enabling the creation of new public value \cite{nachit2021digital, ylinen2021incorporating}. There is also a corresponding technical part regarding project management. In what follows, the change management in terms of laws and regulations is best accounted for in the areas that are affected by those regulations and laws, e.g., Data. Consequently, the part of this manuscript specifically devoted to Change Management refers to the project management engineering. 
            
            Section \ref{sec:Proc_Change_Management} is devoted to this topic. }  
            
            \item { {\bf{Frameworks and Maturity Models}}. These models focus on the major technological, inclusive, and organizational elements of which a \emph{PA} is composed, in order to be able to measure theirt performance and establish the progress achieved in the \emph{DT} strategy undertaken  \cite{dener2021govtech, nerima2021towards}.
            
            Section \ref{sec:Proc_Frameworks} is devoted to this topic.} 
            
        \end{itemize}

    \end{itemize}

    \subsection{Two Paradigmatic Examples} \label{sec:Res_Driving_Examples}

    Each Public Organization may have its own \emph{DT} agenda and plan, which may vary according to factors such as geographic location, size, cultural, economic and infrastructural contexts, e.g., \cite{balashov2020prospects, bousdekis2020digital, miranda2021behind, umar2022digital}. Comparative studies also exist, as for instance:   China, Canada and Estonia;  \cite{lopes2018public}; US and UK  \cite{li2018keys}; Australia, Denmark and the Republic of Korea \cite{meyerhoff2020digital}. Estonia is particularly appreciated in terms of \emph{DT} \cite{dener2021govtech, kitsing2019alternative}, to the point of being covered in the general press, e.g., the New Yorker \cite{New_Yorker}, although some criticism is present \cite{kitsing2018janus}.

    Given the above State of the Art, as anticipated and motivated at the beginning of this section, we now introduce two real examples by focusing on their responsiveness aspects: Barcelona \cite{i2017government, monge2022new} and Chicago \cite{mcbride2019does}.

    \begin{itemize}
    
        \item {{\bf{Technological  Responsiveness}}}
        
        \begin{itemize}
            {\item {\bf{{Smart Cities}}}}. Barcelona, thanks to a budget allocation of 1.288 million EUR, has launched three key \emph{DT} initiatives. 
            The first one is the reorganization of data localization, through the establishment of a Municipal Data Office, headed by a Chief Data Officer. 
            The second one is the mapping of the entire Barcelona Data System, integrating each of the existing datasets into a single \emph{data lake} \cite{margetts2013second}, developed for this purpose, according to the Open Standards defined by the World Wide Web Consortium (\emph{W3C}) \cite{w3consortium} and referred to as the City Operating System (\emph{CityOS}).  It is based on \emph{API} and the data within it are now organized and interconnected thanks to the design of a standardized ontology for the city of Barcelona. An additional data-sharing platform, referred to as \emph{Data Exchange},  is connected with the \emph{CityOS} data lake to ensure a continuous two-way flow of data between the City and the World. 
            The third one is the renewal of the open data portal through the \emph{CKAN} tool \cite{CKanDMS}, to ensure that public, private and personal data can be transformed into a new data-driven social infrastructure. It is worth pointing out that the city-wide data governance model of Barcelona is an extension of the open government agenda promoted by several cities around the world \cite{barns2016mine, hawken2020open}, whereby cities support Open Data platforms for civic engagement and improved digital services to address a range of broader challenges, such as the implementation of Smart Cities.
            
            In order to make clear what follows, it is useful to recall the EU DECODE (\textit{Decentralised Citizen Owned Data Ecosystem}) project \cite{DECODEproject}. Its goal is to develop a combination of decentralized software technologies, such as Blockchains and Cryptography, to give citizens more control over access and usage of their data. The DECODE technology allows data to be encoded and shared anonymously. In addition to what mentioned so far, Barcelona has leveraged \emph{DECODE} through the \emph{citizen Science Data Governance} pilot project, which uses environmental sensors, placed inside and outside participants homes, to detect noise and pollution levels. 
            
            A non-trivial part of this project is the level of detail with which these data are visualized. Data from the \emph{IoT} networks of sensors are collected through the open source platform \emph{Sentilo} \cite{SENTILOproject} in such a detailed and specific manner that individual homes can be identified. This raised concerns about the privacy of this data, as homeowners feared that its use could result in the profiling of pollution-prone buildings and homes, which would hurt house prices or insurance premiums. With the mentioned \emph{DECODE} pilot project, the focus was on developing rules that would allow users to code and share their data with different target groups and with different specificities, generating more trust in the use of the collected data.
            
            Chicago, through the creation of a good quality Open Government Data (\emph{OGD}, for short) portal, continuously improved since 2012, provides data visualization tools on over 550 datasets, a number that continues to grow, and is relevant to the city. Currently, in the available Literature on \emph{OGD}, it is generally pointed out that the \emph{OGD} in the portals that host them are often not accessible, clean, or easy to use \cite{young2017civic}. Remarkably, none of these shortcomings seems to have been reported for the City of Chicago. Indeed, from the responses acquired through interviews, several interviewees were appreciative of the availability and quality of the \emph{OGD} that are available on the Chicago \emph{OGD} portal  \cite{Chicago_OGD_Data}. This goal was achieved through a careful processing pipeline in which data were extracted from data owners, e.g. \emph{PAs}, cleaned and transformed through data cleaning techniques, and uploaded periodically to the \emph{OGD} portal. Approximately 99\% of the data in the \emph{OGD} portal follows this processing pipeline. Maintaining the data quality levels present on the \emph{OGD} portal requires great citizen participation, and an active engagement of the \emph{PAs} that are owners and providers of this data \cite{nikiforova2021open}. With this initiative, Chicago is becoming a reference model of increasing sensitivity to data, which is useful for the creation of digital services, following a paradigm that is more and more open and collaborative, and less and less driven by top-down approaches \cite{nikiforova2021open}. 
            Another important example of Smart Cities in Chicago is in the reduction of the exposure of citizens to foodborne diseases \cite{Chicago_foodborne_illness}. The City of Chicago, in collaboration with its Department of Quantitative Research and Analysis of Allstate, has developed a predictive machine learning model that takes into account various data sources, such as waste, crime, and sanitation data, to support the numerically small staff of the Chicago Department of Public Health ({\emph{CDPH}}) in prioritizing food inspections to be carried out \cite{Chicago_FI_ML}. The model works by ranking restaurants by the probability that they have a critical food safety violation. The head of the \emph{CDPH}, through a simple Shiny web application \cite{Shiny_apps}, is able to assign food inspectors first to the highest-risk restaurants. By using this model, potential foodborne illnesses could be prevented or their severity limited, as the violations were identified and treated earlier with respect to what would have been possible with previous selection methods.

            {\item {\bf{{\emph{Data Governance}}}}. In addition to the aspects of data governance regarding Smart Cities, Barcelona has adopted a series of new standards, technologies, and practices, which have inevitably enabled new ways of managing data by different stakeholders \cite{i2017government, monge2022new}, with the result of increasing transparency, simplicity and objectivity, thereby providing a route to technological and data sovereignty.  This has been achieved through the appropriate use of procurement clauses, e.g., contracts. The interested reader can find the Barcelona ICT Procurement Guide in \cite{barcelona2018procurement}. In particular, and in regard to  data, a minimum set of requirements are mentioned in regard to availability, accessibility, privacy-compliance, and shareability as Open Data among the various City Departments. In particular, they ensure that decisions around who produces, owns and exploits the data generated in the City remain in public hands. Those procurement guidelines are useful case studies for other Cities \cite{monge2022new}. 
            Although Barcelona is a success case in this area, it is to be mentioned that innovative and effective public procurements involving digital systems in the \emph{PA} may be challenging \cite{farshchian2020experiences}.  
                
            It is not sufficiently clear how some aspects of data governance have been handled in Chicago, but in the blog of the Open Data Portal Development Team of the City          \cite{Chicago_OGD_Team}, the process of data collection and accountability is documented specifically for the different types of data collected. The City of Chicago prioritizes personal privacy in the development of datasets for publication. For example, for the Taxi and Transportation Network Provider Trips (TNP or "ride-share") datasets, an anonymization and aggregation technique has been designed and implemented to reduce the risk of passenger re-identification, while enabling favourable public use of the data (see \cite{Chicago_Taxi_TNP} for further details). }
   
        \end{itemize}

        \item {{\bf{Inclusive Responsiveness}}}
        
        \begin{itemize}
            {\item {\bf{Skills and Co-Creation}}. In October 2016, the Barcelona City Council, with an allocation of 75 million EUR to be spent annually on \emph{DT}, planned to provide public services through an approach based on free software, Open Data sovereignty, and the adoption of Agile development methods, as discussed in \cite{i2017government}.
            
            The main challenges addressed in their \emph{DT} plans give rise to several initiatives as follows. First, the launch of an educational programme (\emph{Steam Barcelona}), focusing on building competencies within city organizations, with the aim of strengthening the digital skills of the citizens. Second, the combined utilization of iterative and Agile development methods, for reducing the burden on citizens to use services (\emph{City empowerment}). Third, the design and deployment of new guidelines on the design and accessibility of public services.
            
            With reference to \cite{mcbride2019does}, regarding the City of Chicago, the relationship between \emph{OGD} and co-creation is addressed, in relation to factors that play a role in the co-creation  component of \emph{OGD}-driven public services. The result is the identification of a set of key factors for \emph{OGD}-driven co-creation. Specifically: motivated stakeholders, innovative leaders, proper communication, existing \emph{OGD} portal, external funding, and Agile development.  The interested reader is referred to \cite{mcbride2019does} for further details regarding those factors, since we limit ourselves to discuss Agile development within the {\bf Organizational Responsiveness} below.

            There are also some lessons to be learned from this study. In fact, the authors also reported the main barriers to the publication and reuse of \emph{OGD}, such as the widespread lack of understanding of \emph{OGD} and their benefits. One of the main challenges to the co-creation of public services is the need to redefine the roles of public and private actors in the public service creation process. Some other barriers are connected to the figure of the citizen, such as the internal motivation of participants, personal characteristics, awareness of participation opportunities and participatory skills, perceived ability to participate in co-creation initiatives, trust in co-creation initiatives, the relative importance of the service to be co-created and mutual trust between Government and citizens. }
        
            {\item {\bf{Leadership}}. There are many facets to this topic. Barcelona exemplifies one of them. Specifically, the establishment of a managerial figure such as the Chief Technology and Digital Innovation Officer, to support the city's administration, thanks to which a series of politically and managerially strong reforms could be initiated \cite{monge2022new}. Chicago exemplifies another one. Specifically: technologies, e.g., data analysis techniques that allow better leadership because they support decision-making processes, aiding managers in exploring and solving some of the most difficult problems facing the city \cite{mcbride2019does}.}
            
        \end{itemize}

        \item {{\bf{Organizational Responsiveness}}}
        
        \begin{itemize}
            {\item {\bf{Change Management}.} The City of Barcelona, in 2017, within its \emph{DT} transformation plans,  has provided guidelines for project management that recommend the use of Agile methodologies \cite{i2017government, monge2022new}.
            As a matter of fact, Barcelona has developed its own Agile methodology as a variation of the SCRUM Framework, referred to as SCRUM$@$IMI since the Institut Municipal d'Inform\`atica has had a major role in the adaptation of SCRUM to the  Barcelona ICT needs. The interested reader can find a detailed account of this initiative at \cite{barcelona2017agile}. 

            As for Chicago, in terms of Agile development \cite{mcbride2019does}, according to the opinion of several interviewed stakeholders involved in the development of many projects, although the implementation of services did not explicitly follow Agile development methodologies, many of the characteristics of such approaches were however present in the development of services. The interviewees have emphasized some of these characteristics, considering them crucial to the success of the project, in the design, implementation, and service delivery phases. Namely, speed of development; release of a minimum viable product (\emph{MVP}); validated learning; incremental development; constant testing; and the ability to respond quickly to feedback and evaluations.

            {\item {\bf{Frameworks and Maturity Models}}. The Barcelona City Council has continuously collected feedback and, in terms of metrics, measured various performance indicators on the services provided in order to monitor signs of progress on the expected results of the adopted \emph{DT} strategy \cite{i2017government}.} It is not clear how progress on the expected outcomes of the initiatives implemented in the City of Chicago is measured, as we found no authoritative documents on this topic.}
                
        \end{itemize}        
        
    \end{itemize}

\section{Data} \label{sec:Res_Data}

    We discuss here, in detail, the data ecosystem.

    \subsection{A Glossary of the Data Ecosystem} \label{sec:Res_Data_Ecosystem_Glossary}

   For the convenience of the reader, we describe the following well known  general terms: \emph{Open Data}, \emph{Linked Data}, and \emph{Linked Open Data}.
    
    \begin{itemize}
        \item {\emph{Open Data} are accessible, exploitable, modifiable, shareable by anyone for any purpose, including commercial purposes, and released under an open license \cite{Open_Data}.}
        
        \item {\emph{Linked Data} are structured in such a way as to be interconnected with other data sources to become more useful, promoting discoverability and interoperability. They are built on standard Web technologies such as \emph{HTTP}, \emph{RDF}, and \emph{URI}, but instead of using them only to serve web pages to human readers, they are as well used to share information in a machine-readable way \cite{Linked_Data}.  This type of data has evolved to encode and model knowledge coming from different sources. A notable example are the \emph{RDF Knowledge Graphs} \cite{RDF_Knowledge_Graphs} and related ontologies, built on \emph{Open Data}, that formally model domains of interest. } 
        
        \item {\emph{Linked Open Data} are the intersection of the previous two categories.}
    \end{itemize}

\begin{figure}[!ht]
        \centering
        \includegraphics[scale=0.4]{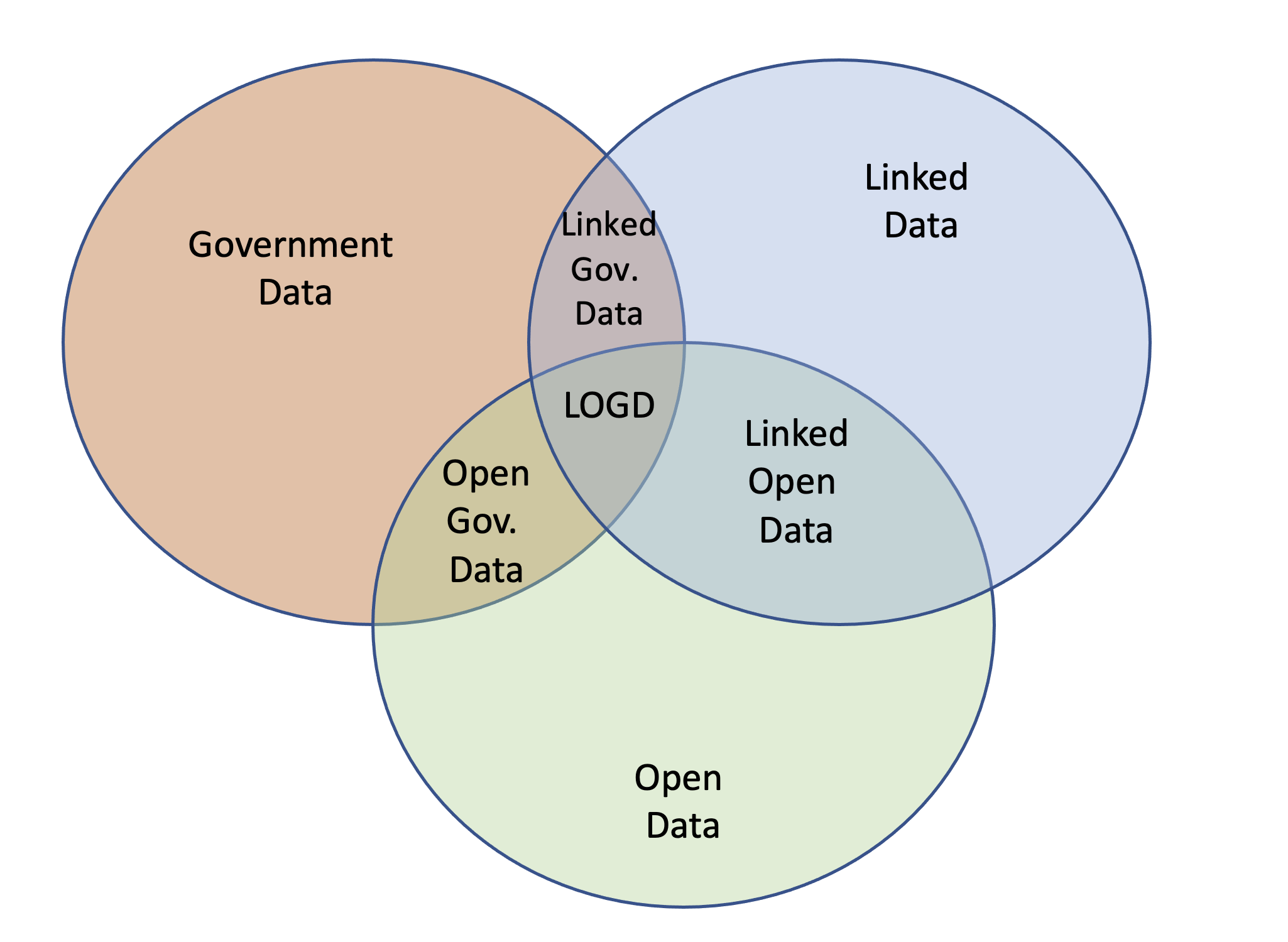}
        \caption{Relationships between Government, Open, and Linked Data (LOGD is Linked Open Government Data). Adapted from \cite{attard2015systematic}.}
        \label{fig:OD_OGD_LOGD}
    \end{figure}

   Government data is any information, in any form, that is created or obtained by the Government in the course of its business. When the data are public, we distinguish \emph{Open Government Data} (\emph{OGD}), \emph{Linked Government Data}, and \emph{Linked Open Government Data} (\emph{LOGD}), respectively. Figure \ref{fig:OD_OGD_LOGD} represents the relationships  among \emph{Government Data}, \emph{Open Data}, and \emph{Linked Data}  \cite{attard2015systematic}. (Linked) Open Government Data make easier for citizens, researchers, developers, and businesses to access and use the data to create new applications, analyze social and policy trends, and develop transparency and accountability.
    
\subsection{Collection and Management of Open Data: \emph{PA}s Are Special } \label{sec:Res_Data_Ecosystem_collection}
   Many of the data that we have described are represented using technologies related to the \emph{Semantic Web} \cite{Semantic_Web}, via Open Data standards, which are defined by the W3C and supported by most technology providers, especially those offering data management tools. However, the collection and management of Open Data by a \emph{PA} seems to require innovation in the processes adopted to carry out  those  tasks \cite{zeleti2019agile}.  In the mentioned study, such an  innovation is characterized in terms of Agile methodologies: the ability of an organization to capture emerging needs and promptly associate them to the current data processes,  in order to obtain innovative data-driven products and services. Based on the empirical study of four \emph{PAs}, a process model for the achievement of  the mentioned agility, is proposed (see \cite{zeleti2019agile} for details). 
   
   Improving the usability of some collected data also requires innovation. Indeed,  as discussed in \cite{osagie2017usability}, although there is a proliferation of Open Data platforms,  their usability for the non-specialist is perceived as poor, mainly due to the fact that they have been designed by software specialists. The mentioned study also provides the evaluation of the usability of an  Open Data platform by non-specialists, for the City of Dublin, pointing out the need for innovative designs of user interfaces. Open Data integration is also a serious obstacle to their fruitful use, with some proposals on how to overcome them, accounting for implementation strategies and organizational models \cite{austin2018path}.

    \subsection{From a Data Ecosystem Abstraction to its Concrete Realization: Some Examples} \label{sec:Res_Ex_DataEcosystem}
    
    Figure \ref{fig:OD_OGD_LOGD} provides a general description of a data ecosystem, which can then be realised in several ways, that should be compliant with the Open Standards defined by the W3C \cite{w3consortium}. Two incarnations of a data ecosystem have already been presented and discussed in Section \ref{sec:Res_Driving_Examples}: the Barcelona CityOS data lake and the Chicago Open Data portal. We now provide three additional examples, concentrating on their technical aspects and pointing out their usefulness. 

    \begin{itemize}
        \item {\textbf{Open Data Catalogs for the PA}. They are software applications that build inventories of data resources of a given \emph{PA},  in order to help data professionals and stakeholders to find relevant data for analysis-related uses \cite{kuvcera2013open}. They are based on metadata, which provide additional data/information about data resources. The intent  is  to help catalogue users to understand the structure, nature and context of the data available in computer systems and decide whether they are suitable for their needs.  One of the earliest relevant examples of an Open Data Catalog proposal is the \emph{OGD} Catalog of the Czech Republic: it serves as a single access point to the \emph{OGD} datasets, supporting the discoverability and reusability of the available \emph{OGDs}. Another, more recent example,  is provided in a case study of the Italian \emph{PA} \cite{cordella2019government}, conducted during the period April-December 2017, in which the use of the \emph{OGDs} favour the implementation and integration of services (or digital platforms) such as \emph{pagoPA} (payments system to \emph{PAs} and public service providers in Italy), \emph{SPID} (Italian Public Digital Identity System), and \emph{ANPR} (Italian Register of Resident Population). These new services are based on \emph{OGD} Catalogs, and in addition to the use of Open Standards, and Open Software, are designed as modular structures, which facilitate their evolution and reduce the complexity of coordination between the actors involved in the co-creation processes of the various projects, as also as reported in \cite{legner2017digitalization, wolff2019will}. 
               
        For a successful integration of services, a good interoperability framework of information sources must be provided.
        In general, it organises the exchange of data and interoperability between different services, data centers and \emph{PAs}. It consists of a set of specific design rules, documents and toolkits for software developers (e.g. Software Development Toolkits). 

        For the specific case study, the main part of the interoperability framework is the Data Analytics Framework (DAF), which collects and processes data from \emph{PAs} and external actors to make them publicly available and accessible through a Web user interface, and defines protocols and regulations that facilitate the integration and orchestration of services. The DAF empowers each \emph{PA} to orchestrate the creation of public value by establishing the actors that can have access to the data and the terms under which they can access them. Uploaded data are supervised by the Data Protection Authority \cite{Data_Authority_IT}, which safeguards the privacy of citizens and evaluates how other public agencies use their data. Therefore, each level of government and the different public agencies are responsible for regulating how data is accessed, according to their administrative and political responsibilities. 
               
        Specifically, a \emph{PA}, as well as a private company, can make data available to the public through the DAF, and can also indicate  who can access that data and the ecosystem on which that data should operate. As expectations and needs change, data access settings can be modified to adapt to emerging needs and requirements. When public agencies upload their data to the DAF, they fill out a privacy form to ensure that the data is privacy compliant, so as to avoid any negative effects on citizen privacy.  }

        \item {\textbf{Cloud-based Open Data Federation: the CLIPS experience} \cite{filograna2016cloudification}. CLIPS is a cloud-based approach for migrating public services to the Cloud, based on the use of microservices. It involves four European Cities: Bremerhaven (Germany), Lecce (Italy), Novisad (Serbia) and  Santander (Spain). It is based on the \emph{Open Data} because, in addition to being a useful resource for developing new value-added services, they seem to be valuable for  exploring potential transnational business opportunities. The CLIPS platform includes an \emph{Open Data Federation} node to allow access to the \emph{Open Data} sets from different federated Municipalities, as if they were a single data source for front-end applications. The main innovation of CLIPS is to provide a usable methodology, that enables Government employees and other external stakeholders to collaborate on new projects and service delivery from a set of basic building blocks, available in the Cloud. This offers the ability to respond more quickly, reduce service delivery costs and be more responsive to end-user needs.
        It defines an approach for building an ecosystem in which \emph{PAs}, small and medium-size enterprises, and citizens can co-create new and innovative public utility services. 
        The CLIPS platform is designed as a three-tier cloud platform, including: an Infrastructure-as-a-service (IaaS) which includes all the required modules to provide basic cloud resources like computation, storage and networking; an application serving and development functionality of traditional Platform-as-a-service (PaaS); and a convenient marketplace for the developed cloud-based services and microservices, typical of a Software-as-a-service (SaaS).
        It consists of several modules, such as authorization, authentication, and monitoring of data access, as well as providing an \emph{API} to connect the microservices present with each other.  
        Data security aspects are also addressed. In fact, the CLIPS Security strategy, in addition to common security best practices (e.g., ISO/IEC 27002, ISO/IEC 27017, ISO/IEC 27018), is to adopt some innovative techniques and approaches from the open-source community as well as from other European Projects, such as "Secure idenTity acrOss boRders linKed" (\emph{STORK}) \cite{STORK_project}, enabling citizens to use their national credentials in \emph{PA} applications provided by foreign States and to securely transfer their sensitive data between the States. }

        \item {\textbf{Data and Smart Cities}.  Smart Cities are perceived as data engines \cite{lammel2020metadata, moustaka2018systematic}, e.g.,  IoT infrastructures, social networks, wearable devices, etc. generate valuable data that can be used to improve or offer new services to the citizens. However, 
        due to its volume and heterogeneity, the collection of data produced by Smart Cities (including the creation of related metadata) requires a non-trivial effort in the verification of its correctness and quality \cite{lammel2020metadata}. Metadata can describe different information sources and can be collected and catalogued within appropriate Open Data Catalogs, such as the open-source solution already mentioned CKAN. Moreover, metadata can be represented through data vocabularies designed to facilitate interoperability between open data sources available on the Web, e.g., using the DCAT-AP metadata profile \cite{dcat-ap}. Fundamental turns out to be the implementation of a set of guidelines and documents such as API documentation, and planning documents, systematically discussed and agreed upon with public officials responsible for providing datasets to improve discoverability, understandability, and further processing of data.
        As shown in \cite{habibzadeh2019smart}, the implementation of these solutions involves a careful design phase of the technology infrastructure (cloud/edge) related to Smart Cities, with an emphasis on the data acquisition plan. The infrastructure  must be the pillar of processing and  storage of data  and also include data analytic tools and methods  finalized to  the implementation of   robust machine intelligence solutions available to the city government for the benefit of its citizens. To this end, 
        three distinct taxonomies of data analytic tools serving Smart Cities  are proposed in \cite{moustaka2018systematic} and referred to as the  DMS Taxonomy, i.e., data, methods and services. }    
        
    \end{itemize}

\section{Technology}\label{sec:Res_Technology}

    We discuss here, in detail, the two main technical aspects we have considered concerning {\bf{\emph{Technology}}}: Smart Cities and data governance. As anticipated in Section  \ref{sec:Tour_Responsiveness}, for Smart Cities, we concentrate on Digital Twins.

    \subsection{Smart Cities: Digital Twins} \label{sec:Tech_SC_IoT}
    
     Being specific to Smart Cities and following \cite{DWZ2021}, the major characteristics of Digital Twins are: accurate City Mapping, for instance, of roads and public illumination; interaction between the virtual and real \vir{objects}, e.g., people and their \vir{avatars}; software definition, e.g., platforms that simulate the real city in a virtual space; intelligent feedback, e.g., evaluation of the effects of city plans and initiatives before realization.  Interestingly, it has been argued that their realization may enable an acceleration of NetZero emissions in government critical infrastructures \cite{reeves2022digital}. A further refinement of the technical characteristics of  Digital Twins is proposed in \cite{zhang2022framework}, although its major contribution seems to be the account of Digital Twins initiatives in China, USA, and France. 

    Although there are many national and city initiatives regarding Digital Twins, e.g., \cite{Data61, NSW_DT, Virtual_Singapore,zhang2022framework}, we have found only a limited number of academic papers covering the subject. One is in regard to cross-border Smart Cities, i.e., Helsinki and Tallinn. Recalling that an urban operating system is a network of sensors that can acquire data regarding the city which, in turn, can be transformed into  \vir{knowledge} \cite{soe2017smart} and pointing out that the X-Road data infrastructure \cite{saputro2020prerequisites} is one of the pillars of Estonian \emph{DT}, a cross-border urban operating system involving both cities is proposed in \cite{soe2017finest}. The intent is to  have an integration of the \emph{DT} that is involving only each of the mentioned cities. For completeness as well as relevancy for this Tutorial, we mention that the notion of urban operating system is investigated in depth in \cite{luque2020urban}, with various examples of it. The study points out the modest impact that it may have on city planning and its contradictions. 

    Overall, Digital Twins have the potential to bring significant benefits to Smart Cities, including  better evaluation of city plans, and potentially even achieving NetZero emissions, but further research is needed to fully understand their impact.

    \subsection{Data Governance}\label{sec:Tech_CyberSec}  
    
    In the context of our Tutorial, we focus on two particular aspects of data governance, i.e., privacy and cyber-security. \\
    
    In terms of privacy, in the international scenario, there are several National Data Protection Authorities. An exhaustive list of these Authorities can be found through an interactive map on the website of the French National Commission for Information Technology and Civil Liberties (CNIL) \cite{CNIL_map}. The main function of the individual Authorities is to protect the privacy of citizens and assess how the \emph{PAs} (or other organizations) use their data, for example, by keeping under control the data they publish on their respective institutional web portals. Barcelona is a good example in terms of control of the data, regarding availability and detail of access, as discussed in Section \ref{sec:Res_Driving_Examples}. As well argued in \cite{potoczny2019encrypted}, the amount of data that is collected within Smart Cities initiatives, once made public, even in an anonymous form, can be subject to cross-reference attacks that could capture private information. In order to address this problem, the mentioned paper proposes solutions and use-cases. Interestingly, that study is a pilot project funded by the U.S. Department of Homeland Security that has the intent to demonstrate how data privacy technologies can be of help. 
    
    In terms of cyber-security, it is well known that there is a proliferation of Cyber-Security Agencies, e.g., the European Union Agency for Cyber-Security (\emph{ENISA}) \cite{ENISA}. This is not surprising, given the increase in the number and quality of the attacks of which we have news in the past few years \cite{clusit2022report}. However, since \emph{PAs} are also the object of those attacks, it is surprising that there is only a limited number of papers that have emerged from the literature regarding \emph{DT} that address cyber-security issues, as we outline next. 

     With regards to data, in \cite{bounabatgovernment}, the Organisation for Economic Co-operation and Development (OECD) \cite{OECD} recommends maintaining a strong balance between the need to provide timely official data and the need to provide reliable data, as well as to manage the risks associated with the increased availability of data in open formats and those related to digital security and privacy. A related issue is the design and management of government data centers architectures, in particular regarding security. Indeed, those centers, due to the heterogeneous nature of services they offer and software they host, are vulnerable from the point of view of security. A proposal on how to achieve ISO/IEC 27000 security standard, a model of government data centers architecture, has been proposed in \cite{urbanczyk2021application}. More in general, as discussed in \cite{almeida2022tecseg}, there are several initiatives in many countries having the goal to provide methodologies for security assessment. This latter consists of evaluating an information system from the attacker's point of view, with the aim to provide a systematic review of weaknesses in information systems,  with a corresponding assignment of probabilities of attack via each weakness, offering also a scale of severity levels of damages. Recommendations for corrections are also offered.

     Smart Cities and their associated technologies, being relatively novel, are also object of study in terms of security. A  specific analysis regarding IoT devices and related technological infrastructures, is given in \cite{saba2022toward} (but see also \cite{zheng2020towards} and references therein). Indeed, due to their interconnected nature, IoT technologies make data security a more complex challenge with respect to the past. Therefore, ensuring the security of IoT products and services has become a top priority.  To this end, an entire framework, referred to as SAO, regarding the automation of IoT security has been proposed in \cite{zheng2020towards}. It has the merit of being grounded on a recent review of the State of the Art, clearly describing challenges and proposing solutions. SAO integrates the key elements for security automation and orchestration for IoT systems, including threat modeling, security and privacy by design, trust management, security configuration, threat monitoring, patching, compliance check, and secure data sharing. Another specific analysis is provided in \cite{alcaraz2022digital}, regarding Digital Twins. Indeed, the confluence of a broad set of technologies, ranging from cyber-physical systems to artificial intelligence, and the implicit interaction with the real objects modeled by the Digital Twin, poses new security threats. The mentioned paper offers a classification of them, together with security recommendations on how to address them, via a paradigm that classifies the threats  based on the functionality levels composing a Digital Twin.

\section{People}\label{sec:Res_People}
    
    We discuss here, in detail, the technical aspect we have found concerning {\bf{\emph{People}}} that deserves further attention, based on the Literature  search: \emph{Skills and Co-Creation}.  

    \subsection{Skills and Co-Creation} \label{sec:People_Skills_Cocreation}

    A successful \emph{DT} process requires users not only to acquire new skills but also to know how to interact effectively with them \cite{wolff2019will}. Those skills required to handle \emph{DT} do not only relate to a particular discipline but require a multidisciplinary approach, where the importance of knowing the specific competency levels of the individuals that are part of an organisation and the know-how of the entire organisation itself is recognised as a fundamental requirement. The lack of a coherent educational approach to the acquisition of appropriate skills also hurts e-government users, which could generate problems in the usability of the \emph{PAs} digital services \cite{battisti2020digital}. In \cite{wolff2019will}, the authors, as a possible solution to this shortcoming, propose an educational framework composed of five basic components designed, developed, and tested to achieve the educational goals necessary for a successful \emph{DT} strategy. The components of the framework were intended to define: (a) a competency model useful to describe the required competencies; (b) an educational approach that can be provided by the professional or academic context; (c) a maturity model to monitor progress in the process of acquisition of the required competencies; (d) an appropriate didactic model that is tailored to digital capabilities and demands is essential in order to make competence delivery successful and efficient; and (e) a competency certification system to coach organizations and citizens to understand and communicate their competencies, ensuring transparency and quality. 

    As for co-creation, it is useful to recall from the previous sections that the ultimate goal of a digital \emph{PA} is the co-design and deployment of services that are perceived as being of \vir{value} ( see {\bf {\emph{People}}} in Section \ref{sec:Panoramic_Building_Blocks}). 
    Accordingly, how to achieve that goal and with which methodologies and supporting technologies is an emerging area of research \cite{panagiotopoulos2019public}, that we outline next. 

    In its simplest and easiest to realize form, a co-creation methodology is limited to the participation of a strictly selected set of users, particularly in the initial phase of the creation process, and to the related measurement of their perceived satisfaction degree, through constant feedback collection \cite{vestues2021usingProcess}. However, the intent is to have co-creation methodologies that can handle millions of users, i.e., citizens. It is natural, then, that the IT platforms supporting the \emph{PA} must support those \vir{in the Large} co-creation methodologies. 
    Recalling from \cite{cordella2019government}  that Government as a Platform (GaaP) is a new way of building digital public services using a collaborative development model by a community of partners, providers and citizens to share and enhance digital public processes and capabilities, or to extend them for the benefit of society, its realizations seem to be   
    designed to achieve efficiency. However,
    according to the mentioned study, the efficiency granted by GaaP does not necessarily imply the creation of value for the citizens, a point also made in \cite{lopes2018business, vaira2022smart}. Indeed, as discussed in the mentioned paper, the key to the creation of public value seems to be the modularity of the platform configuration and the ability to consistently coordinate different ecosystems that support public agencies. To this end, a few examples are provided, borrowed from the private sector and involving IT giants such as Apple, Google and Amazon. Here we limit ourselves to mention the Apple iOS Support Service \cite{eaton2015distributed}, which enables multiple ecosystems, different in nature,  to interact and coexist. According to the analysis reported in \cite{cordella2019government}, the adoption of analogous models would allow the co-creation (\emph{PAs} and citizens) of value services with a \vir{large scale} involvement of active actors. The importance of adequate digital platforms for the co-creation of value involving a large number of actors is also identified as a key success factor in \cite{vestues2021usingProcess}. 
    A paradigm shift from crowd-sourcing and social media monitoring to IoT has also been proposed, with a pilot project that has been set-up in a Municipality in Sweden \cite{hedestig2018co}. 
    
    In addition to what we have mentioned so far, the notion of participatory design, e.g., the involvement of citizens in urban planning, is being analyzed in view of \emph{DT}. A historic account of how that notion has changed over the decades and how it fits a modern view of \emph{DT} is provided in \cite{pilemalm2018participatory}. An important related topic is the co-creation of integrated public services. That is, ideally, a one-stop platform for the citizens that integrates the available services to them. The State of the Art, mostly regarding EU, is well presented in \cite{tambouris2021towards}.     

    For completeness, we also mention that, in terms of \emph{PAs} and co-creation of value, the Italian public administration as a platform is studied in \cite{cordella2019government}; the Norwegian Labour and Welfare Department is studied in \cite{vestues2021usingProcess}, while a platform supporting co-creation at different levels of governance in Portugal is presented in \cite{sousa2018liberopinion}. A specific platform for co-creation in the area of Urban Planning and in support of previous initiatives, i.e., the International Laboratory of Architecture and Urban Design, has been proposed in \cite{guerreri2021evolving}. Finally, a model based on Digital Twins that allows co-creation, as well as evaluation of the final result regarding public services has been proposed in \cite{petrova2019methodological}, with a planned test of the model in Sofia.

\section{Process}\label{sec:Res_Process}
    
    We discuss here, in detail, the two main technical aspects we have found concerning {\bf{\emph{Process}}}: \emph{Change Management} and \emph{Frameworks and Maturity Models}.

    \subsection{Change Management} \label{sec:Proc_Change_Management}

    As well put in \cite{barcelona2017agile}, although the Agile Manifesto dates back to 2001 and despite the remarkable success that the corresponding methodologies have had in the private sector, their adoption in the \emph{PA} is rather slow. Yet, in the \emph{DT}, Agile  project management methodologies (see \cite{shore2008art}) seem to be the ones that should replace more classic ones, such as Waterfall \cite{WATERFALL}. In order to exemplify this point, the experience reported in \cite{nerurkar2017agile, nerurkar2017analysis} suffices. In the mentioned studies,  the authors point out that the implementation of the e-governance project Digital India Land Records Modernization Program (DILRMP) has highlighted major challenges and complexities, typical of traditional project management. They discuss how an Agile management approach can play a key role in transforming such implementation  from slow and ineffective to  be more responsive, flexible and effective. 

    Documented difficulties in the adoption of Agile methodologies have emerged \cite{dwivedula2020transitioning, mergel2021agile, ribeiro2018acceptance, masombuka2020framework, wipulanusat2019drivers, mohagheghi2021organizational}. The cause is common: the difference in \emph{modus operandi} between the \emph{PAs} and the private sector, resulting in resilience to change, and difficulty in identifying the most appropriate methodologies for the \emph{PA}. Fortunately, studies \cite{reginaldo2020challenges} seem to have identified \vir{agility enables}, i.e., possible actions that can facilitate the transition to Agile models. However, as pointed out in \cite{jovanovic2020agile}, the transition to Agile development models will require the writing of appropriate guidelines to be used to ensure that the development process is Agile. These will depend on the particular requirements of the organization involved in the transition process.

    Although the highlighted difficulties persist, there are many \emph{PA} project management initiatives that use the Agile methodologies, e.g., in the software development Census of the Swedish Government Agencies, the majority of Government Agencies consider their approach to be more agile than planned \cite{borg2018digitalization, lindgren2018digital}. In addition to Barcelona and Chicago, mentioned in Section \ref{sec:Res_Driving_Examples}, the Agile methodologies are applied in several \emph{PAs} \cite{barcelona2017agile, i2017government}, ranging from  National (e.g. UK), Large Cities (e.g. New York), and Regional Governments (e.g. Andalusia). Apart from the above  noteworthy examples, a  systematic and technical presentation of the adoption of Agile methodologies for project management in the \emph{PA} is reported in \cite{aleinikova2020project}. The paper makes also a list of the Agile process automation technologies that are in use, i.e., \emph{SCRUM}, \emph{KANBAN}, and \emph{SAFe} (see again \cite{shore2008art}). A comparison is also performed with classic Waterfall methodologies and it is stated that the Agile ones allow for more transparent projects, effective team building, adaptability to change, lack of hierarchy, lack of bureaucracy, continuous education. Some disadvantages are also reported, such as: the risk of endless product changes; the high dependence on the qualification and experience level of the development team; the difficulty of determining total project costs in a timely manner. Although unclear in its impact, an effort is also made for the identification of the specific characteristics that the Agile methodology should have for its use in the public sector \cite{bogdanova2020agile}. The mentioned paper reinforces the difficulties already mentioned and that must be overcome for such a change of project management. Moreover, it stresses that project management should be reconfigured to provide team autonomy, to some extend. Once again, the barrier being routine practices difficult to abandon and obsolete regulations. 
        
    A more specific evaluation of Agile methodologies in the \emph{PA}, regarding DevOps \cite{DevOps}, is provided in \cite{mubarkoot2021assessment, yarlagadda2018public}, where it is considered how to bring best practices from the production world into \emph{PA}, making the flow of information more fluid. As a result, the adoption of \emph{DevOps} promotes organizational responsiveness, which is useful for improving productivity and performance. At the same time, \emph{DevOps} breaks down organizational barriers by promoting information exchange through the use of shared metrics and feedback mechanisms between development teams, as reported in \cite{forsgren2018devops}. 

    By bringing \emph{DevOps} into the public sector, an effective teamwork and a consequent open flow of knowledge among \emph{PA} employees are expected \cite{meirelles2017brazilian, stirbu2018towards}. There are initiatives in this regard, as for instance the ones of the Brazilian Federal Government, referred to as  Brazilian Public Software (SPB). The objective is to promote sharing and collaboration enabled by Free/Libre/Open Source Software (FLOSS) solutions for \emph{PA} \cite{meirelles2017brazilian}. SPB is an interconnected platform based on different FLOSS tools that provides different solutions for collaborative software development, with the purpose of enabling Brazilian \emph{PAs} to share information, experiences, and best practices about the use of these tools (see \cite{SPBportal} for more details about the architecture and operational manuals).
    
    Furthermore, since  transparency and openness are among the core principles of \emph{DevOps} practices, their use is expected to simplify bureaucracy and decrease corruption in public service delivery.  A punctual analysis regarding  the benefits of using DevOps Process Model in the \emph{PA} is presented in \cite{zazour2020devops}. It involves seven Saudi Arabia \emph{PAs}, evaluated with the use of the Bucena DevOps Maturity Model \cite{bucena2017simplifying}. That study concludes that the use of DevOps is promising although DevOps cultural aspects, process, and technologies need to be strengthened. An additional study proposing DevOps for the generic support of Digital Transformation is presented in \cite{sarah2020devops}, being in agreement with the papers mentioned so far. 
    
    We mention that there are also experiences indicating that classic Waterfall and Agile methodologies can synergically co-exist. Indeed, we learn from  the case study in \cite{mantovani2020characteristics},  involving the development of projects through Agile methodologies of some Brazilian governmental organizations, that although the adoption of such methodologies fosters an improvement in the quality of the public services created, these projects achieve greater success when conducted in combination with other traditional software development approaches. 
   
    Finally, Agile software development in the public sector must be scalable, i.e., able to work for relatively small projects, coming from small realities such as cities, to large national and international projects, for example through the adoption of the SAFe Agile process automation technology, as reported in \cite{fagarasan2022delivery, ghimire2020scaling}. To this end, it is of interest to mention a recent review \cite{edison2021comparing} regarding the use of Agile methodologies on a large scale. Although one would expect that the \emph{PA} would be the area with the most involvement, it is somewhat disappointing to report that only $5\%$ of the initiatives reported there belong to the public sector.

    \subsection{Frameworks and Maturity Models}\label{sec:Proc_Frameworks}
    
    Over the past decade, various frameworks and models have been developed to measure and monitor the degree of digital maturity achieved by Digital Transformation Strategies. To date, however, it is not possible to choose one among them for which any organizational reality can be perfectly modelled, whether private or public. Each of these captures a particular set of indicators and uses different tools to collect information to be used to quantify the indicators. One of the tools  is certainly interviews, with the possible addition of document analysis \cite{anders2019agility, cordella2019government, fangmann2020agile, feher2018digitalization, jonathan2019digital, juniawan2017smart, kane2015strategy, luz2018building, luz2019adopting, mantovani2020characteristics, schedler2019smart, tangi2020barriers, tangi2021digital, vogelsang2019taxonomy, ylinen2021digital, ylinen2021incorporating}. In the mentioned case studies, semi-structured interviews are mainly conducted with various IT professionals from public and private organizations, actively involved in \emph{DT} processes, over different periods in order to measure the degree of digital maturity gained. Several barriers and success factors emerged from the interviews, which are useful for a comprehensive understanding of \emph{DT}. The results show that this survey instrument is quite valid, as effectively reported in \cite{schedler2019smart, tangi2020barriers, tangi2021digital} (see respective Appendix Sections).
   
    In addition to the model specific to interviews,  many general maturity models have been developed over time. For most of them, based on variables specifying the model, the \vir{end-result} is the value of an index that assesses the level of achieved maturity. Some follow macro-economic factors on a national or international level, as in \cite{GI_Index, aniscenko2017regional, dener2021govtech, CAF_Index, EC_eGov_Benchmark_Index, OECD_Index, Dig_Adop_Index, UN_Index, idowu2021framework, jussupova2019digital}. Others, however, refer to micro-economic factors related to individual organizations, as in \cite{bucena2017simplifying, CMMI, nerima2021towards, IGMM, E-Gov-MM}. 
    
    In regard to the first group, we discuss only the GovTech Maturity Index (\emph{GTMI}) developed by the World Bank \cite{dener2021govtech}, as part of their \emph{GovTech} initiative (Government and Technology) \cite{GovTech}, since it appears to be the most exhaustive maturity model currently available. It is worth pointing out that \emph{GovTech} is an approach to the modernization of the public sector, through innovative technological solutions, that promotes a simple, efficient, and transparent Administration with the citizens at the center of the reforms. There are about 80 \emph{GovTech} initiatives worldwide, with good practices observable in 43 countries out of the 198 observed. In this context, \emph{GTMI} is a comprehensive measure of the \emph{DT} in a given country. It is based on 48 key indicators and it is defined to collect data from 198 countries. \emph{GTMI} measures key aspects of four focus areas of the \emph{GovTech} initiative: supporting Core Government Systems (\emph{CGSI}, 15 indicators), improving Service Delivery (\emph{PSDI}, 6 indicators), Engaging citizens (\emph{CEI}, 12 indicators), and promoting the Enabling factors of the \emph{GovTech} initiative, such as building digital skills in the public sector and an environment conducive to innovation in the public sector (\emph{GTEI}, 15 indicators). Each of the indicators is associated with a certain score and a certain weight, the latter based on the opinions of some domain experts on the relative importance of the selected indicator. Using these scores and weights, the \emph{CGSI}, \emph{PSDI}, \emph{CEI}, and \emph{GTEI} scores are calculated. The final \emph{GTMI} score, on a $[0,1]$ scale, is calculated as the arithmetic mean of the four scores just mentioned. See \cite{dener2021govtech} for more explanatory details on the indicators. All 198 countries were grouped into four categories: from A (leaders in \emph{GovTech}) to D (minimal attention in \emph{GovTech}) according to their \emph{GTMI} score.  
    
\ignore{
    Specifically:  
    \begin{enumerate}[label=\Alph*.]
        \item {Governments leaders demonstrating advanced or innovative solutions and good practice in all four focus areas ($0.75 \leq \emph{GTMI} \leq 1.00$)}
        \item {Governments with significant \emph{GovTech} investments and good practices in most areas of interest ($0.50 \leq \emph{GTMI} \leq 0.74$)}
        \item {Governments with ongoing activities to improve some of the \emph{GovTech} focus areas ($0.25 \leq \emph{GTMI} \leq 0.49$)}
        \item {Governments with minimal focus on \emph{GovTech} initiatives ($0 \leq \emph{GTMI} \leq 0.24$)}
    \end{enumerate}
}

    Based on analyses comparing the \emph{GTMI} with other relevant indices, the \emph{GTMI} indicators were found to be consistent and robust, even concerning the analysis of lesser-known dimensions related to particular characteristics of a given Government. Results and good practices presented in \cite{dener2021govtech} demonstrate how the \emph{GovTech} focus areas identified by the World Bank are highly relevant to the \emph{DT} agenda in most countries.
    
    As for the  second group of models, which relates to the micro-economic factors of individual organizations. For conciseness, we will only briefly discuss the Digital Maturity Balance Model \cite{nerima2021towards}. It is oriented towards \emph{PAs} and is based on two axes: digital maturity and importance ratio. The focus is on measuring the balance between the two. Each maturity dimension is assessed by taking into account the importance ratio of this dimension in the Organization. The main categories of maturity dimensions involved are data, IT governance, strategy, organisation, and process. The construction of the model essentially consists of three steps. First, a method must be defined to assess digital maturity. Secondly, a method must be defined   to measure the importance of each dimension of digital maturity pertaining to each of the categories involved. Third, a self-assessment tool must be provided that combines the methods just mentioned, e.g., in the form of an online questionnaire, in which the questions allow the assessment of the digital maturity criteria and the digital relationship attributes. Results show that the use of the model and of the self-assessment tool is  useful and relevant, but needs further refinement to fully correspond to the reality of a given \emph{PA}. 
    
    Interestingly, micro-economic maturity indexes may be of use in measuring other aspects of \emph{DT}, far from the ones they have been designed for. By way of example, the CMMI index \cite{CMMI} has been adapted in \cite{teixeira2020maturity} in order to measure the success  of the adoption of the Agile DevOps methodology in the \emph{PA} project management.

\section{Future Directions }\label{sec:Future_Directions}
   
    \begin{itemize}
        \item { \textbf{Data}.} From what has been discussed in Section \ref{sec:Res_Data}, it is evident that data innovations come from using Open Semantic Web standards in the context of \emph{PA} to represent their information assets. The introduction and use of Open Data is certainly a big step forward since the  advanced functionalities they make available have transformed the \emph{OGD} landscape \cite{charalabidis2018multiple}. Apparently, little attention has been dedicated to the \emph{LOGD}, in particular, to all those activities related to the production and maintenance of quality levels, which facilitate interoperability with other data sources \cite{attard2015systematic}, according to the Open Government principles \cite{janssen2012benefits, toots2017open}. In particular, a domain that needs attention for the \emph{DT} is the one regarding the use of \emph{RDF Knowledge Graphs}, since their use would facilitate the discovery of new data sources and improve their interoperability among different \emph{PAs}. Another aspect that needs to be developed is to set-up mechanisms that strengthen the trust among citizens and \emph{PAs} regarding the use of the collected  data \cite{vaira2022smart}. A related topic is security, in particular regarding the creation of a system of protection balancing the needs of \emph{PAs} and the risks connected to open data and interoperability. 
        Another issue that needs some care during a \emph{DT} is the efficient storage of the vast amount of data that is continuously increasing. To this end,  there is extensive experimental research addressing compression of RDF data in \cite{besta2018survey, brisaboa2020space, liu2018graph, maneth2015survey, MartPrieto2018}, but we found no mention of those techniques as being currently used in the context of \emph{PA}.

        \item { \textbf{Technology}.} As outlined in Section \ref{sec:guidedtour}, Cloud Computing is a main component of any \emph{DT}. Moreover, as discussed in  Section \ref{sec:Tech_SC_IoT}, the diffusion of Smart Cities and Digital Twins are very promising. Somewhat unfortunately, the complexities related to their full scale realization  are far from being addressed and resolved. The difficulties of scaling are best exemplified by a study regarding energy consumption optimization of \vir{only} sixteen buildings in Rome, via Digital Twins \cite{agostinelli2021cyber}. A recent review clearly outlines the five major challenges that need to be addressed \cite{wang2023digital}. Not surprisingly, they range from data collection, storage and analysis  to computing power. Although some research directions are also mentioned, they lack specificity and a clear assessment of how the scale of what has to be managed via Digital Twins affects costs: a city, even a major one, may not be able to economically sustain its full fledged Digital Twin.
        
        Concerning privacy and security, the  adoption of recognized standards, such as ISO/IEC 27001 is strongly recommended, as indicated in \cite{filograna2016cloudification, urbanczyk2021application}. To this end, it is suggested that a more collaborative approach be taken to support security in developing effective and appropriate solutions to security challenges, including increased efforts on technologies \emph{IoT} \cite{saba2022toward}, to prevent attacks or minimize their effects.
        At the State of the Art, there are no evident documented outcomes in the Literature on how these recommendations have been understood and pursued by the \emph{PAs}. The actions, however, appear to be in place, as shown in the timelines of the \emph{NRRP} Plans, i.e., \cite{NRRP_IT}.

        \item { \textbf{People}.} One of the major problems that emerge in  terms of digital skills is the necessity of proper educational efforts, such as courses and tutorials, in particular in developing countries \cite{wolff2019will}. In summary, the development of a  digital education ecosystem is one of the major needs for an effective \emph{DT}. By way of example, actions in this direction are planned in Europe \cite{DigEducActPlan2027} and recommendations are given in the U.S. \cite{AdvDigEquityAll2022}.
        In terms of co-creation, its widespread  adoption within \emph{PAs} requires relevant structural changes, including a sourcing strategy, a governance structure, and a more flexible digital infrastructure, as reported in \cite{vestues2021usingProcess}.

        \item { \textbf{Process}.} It is clear from Sections \ref{sec:guidedtour} and \ref{sec:Res_Process} that the way in which projects are designed, managed and implemented must change in order to achieve an effective \emph{DT}. Agile technologies are one  technical way  of realizing such a change. However, the \emph{DT} is a dynamic process that may generate the need for \vir{new and higher  transformations} that may impact the mission of an organization. For instance, the IT department of a large Finnish municipality, transformed its mission from problem-solving to proactive service delivery, partly through a collaborative approach with business units, as reported in \cite{ylinen2021digital}. Therefore, Agile technologies may well be the \vir{tools}, but a clear plan of what is \emph{DT} is essential. Such a plan and vision may change depending on the scale (local, regional, national), although some coherence among the various levels of the scale must be ensured. To the best of our knowledge, a \emph{DT} approach that accounts for the granularity and hierarchy of the components involved is not present.  
        It is to be said that Agile technologies reinforce the need for capacity-development  of stakeholders \cite{vestues2021usingPeople}, i.e., the acquisitions of digital skills. 
        Moreover, although more collaborative project management approaches are felt as necessary with the goal of interoperability, the lack of agreed processes, the difficulties of interpreting administrative and legislative procedures, and the difficulty of defining authorities and responsibilities are just some of the reasons why interoperability between \emph{PAs} is not achieved, as outlined in \cite{margariti2020assessment}. Again, solutions to this problem are  related to the scale at which we look at \emph{DT}: interoperability may be simple to achieve in a restricted and uniform community and much more difficult in larger and more heterogeneous ones. 
        
    \end{itemize}

\section{Conclusions} \label{sec:Conclusions}

This tutorial presents a guided tour of the main areas of the \emph{DT} of the public sector from the perspective of a computer scientist. We started from an analysis of the literature on Digital Transformation available on some digital libraries well known to computer scientists. Using the query described in Section \ref{sec:methodology} we found almost six thousands of papers related to Digital Transformation, that were reduced to the papers listed in the references that we have used as the basis for this tutorial.
Our study has identified the critical factors of a successful \emph{DT}, and the challenges in the areas of data, technologies, people and processes which have been faced by some public administrations in different countries.
We believe we have given an original synthesis of some problems and their solutions, useful for understanding the main topics underlying efforts of digitally transforming the life of citizens by some public administrations.
Our findings suggest some future directions for research and practice in the four areas mentioned, as discussed in Section \ref{sec:Future_Directions}.


\bibliographystyle{abbrv}

\begin{thebibliography}{100}

\bibitem{kraska2022seattle}
D.~Abadi, A.~Ailamaki, D.~G. Andersen, P.~Bailis, M.~Balazinska, P.~A.
  Bernstein, P.~A. Boncz, S.~Chaudhuri, A.~Cheung, A.~Doan, L.~Dong, M.~J.
  Franklin, J.~Freire, A.~Y. Halevy, J.~M. Hellerstein, S.~Idreos, D.~Kossmann,
  T.~Kraska, S.~Krishnamurthy, V.~Markl, S.~Melnik, T.~Milo, C.~Mohan,
  T.~Neumann, B.~C. Ooi, F.~Ozcan, J.~M. Patel, A.~Pavlo, R.~A. Popa,
  R.~Ramakrishnan, C.~R{\'{e}}, M.~Stonebraker, and D.~Suciu.
\newblock {The Seattle Report on Database Research}.
\newblock {\em Communications of the ACM}, 65(8):72--79, 2022.

\bibitem{abell2021cloud}
T.~Abell, A.~Husar, and L.~May-Ann.
\newblock {Cloud Computing as a Key Enabler for Tech Start-Ups across Asia and
  the Pacific}.
\newblock {Asian Development Bank - Sustainable Development Working Paper
  Series}, 2021.

\bibitem{abied2022adoption}
O.~Abied, O.~Ibrahim, and S.~N. I.~M. Kamal.
\newblock {Adoption of Cloud Computing in E-Government: A Systematic Literature
  Review}.
\newblock {\em Pertanika Journal of Science \& Technology}, 30(1), 2022.

\bibitem{agarwal2017enterprise}
R.~Agarwal, V.~Thakur, and R.~Chauhan.
\newblock {Enterprise Architecture for E-Government}.
\newblock In {\em Proceedings of the 10th International Conference on Theory
  and Practice of Electronic Governance}, pages 47--55, 2017.

\bibitem{agbozo2020towards}
E.~Agbozo and A.~N. Medvedev.
\newblock {Towards a Multi-Channel Service Delivery Model in the Data-Driven
  Public Sector}.
\newblock {\em Business Informatics}, 14(1):41--50, 2020.

\bibitem{agostinelli2021cyber}
S.~Agostinelli, F.~Cumo, G.~Guidi, and C.~Tomazzoli.
\newblock {Cyber-Physical Systems Improving Building Energy Management: Digital
  Twin and Artificial Intelligence}.
\newblock {\em Energies}, 14(8), 2021.

\bibitem{WATERFALL}
M.~Ajam.
\newblock {\em Project Management Beyond Waterfall and Agile}.
\newblock Auerbach Publications, 2018.

\bibitem{al2022healthcare}
J.~Al-Jaroodi, N.~Mohamed, N.~Kesserwan, and I.~Jawhar.
\newblock {Healthcare 4.0 - Managing a Holistic Transformation}.
\newblock In {\em 2022 IEEE International Systems Conference (SysCon)}, pages
  1--8. IEEE, 2022.

\bibitem{alcaraz2022digital}
C.~Alcaraz and J.~Lopez.
\newblock {Digital Twin: A Comprehensive Survey of Security Threats}.
\newblock {\em IEEE Communications Surveys \& Tutorials}, 2022.

\bibitem{aleinikova2020project}
O.~Aleinikova, S.~Kravchenko, V.~Hurochkina, V.~Zvonar, O.~Brechko, and
  Z.~Buryk.
\newblock {Project Management Technologies in Public Administration}.
\newblock {\em Journal of Management Information and Decision Sciences},
  23(5):564--576, 2020.

\bibitem{alghamdi2017desinge}
B.~Alghamdi, L.~E. Potter, and S.~Drew.
\newblock {Desinge and Implementation of Government Cloud Computing
  Requirements: TOGAF}.
\newblock In {\em {2017 11th International Conference on Telecommunication
  Systems Services and Applications (TSSA)}}, pages 1--6. IEEE, 2017.

\bibitem{almeida2022tecseg}
A.~Almeida, J.~P. D.~L. Cassiano, and S.~Ribeiro.
\newblock {TecSEG Project-Research, Development and Innovation in Security
  Assessment Methodologies to Brazil}.
\newblock In {\em {2022 International Conference on Electrical, Computer,
  Communications and Mechatronics Engineering (ICECCME)}}, pages 1--6. IEEE,
  2022.

\bibitem{anders2019agility}
N.~Anders and B.~Schenk.
\newblock {Agility in Public Administration - Is Agility a Possibility and
  Where are its Limits?}
\newblock In {\em Central and Eastern European eDem and eGov Days}, pages
  97--104, 2019.

\bibitem{aniscenko2017regional}
Z.~Aniscenko, A.~Robalino-L{\'o}pez, T.~E. Rodr{\'\i}guez, and B.~E. P{\'e}rez.
\newblock {Regional E-Government Development: Evolution of EGDI in Andean
  Countries}.
\newblock In {\em 2017 Fourth International Conference on eDemocracy \&
  eGovernment (ICEDEG)}, pages 22--31. IEEE, 2017.

\bibitem{arias2018digital}
M.~I. Arias and A.~C.~G. Ma{\c{c}}ada.
\newblock {Digital Government for E-Government Service Quality: a Literature
  Review}.
\newblock In {\em Proceedings of the 11th International Conference on Theory
  and Practice of Electronic Governance}, pages 7--17, 2018.

\bibitem{RDF_Knowledge_Graphs}
H.~Arnaout and S.~Elbassuoni.
\newblock {Effective Searching of RDF Knowledge Graphs}.
\newblock {\em Journal of Web Semantics}, 48:66--84, 2018.

\bibitem{attard2015systematic}
J.~Attard, F.~Orlandi, S.~Scerri, and S.~Auer.
\newblock {A Systematic Review of Open Government Data Initiatives}.
\newblock {\em Government Information Quarterly}, 32(4):399--418, 2015.

\bibitem{Linked_Data}
S.~Auer, V.~Bryl, and S.~Tramp.
\newblock {\em {Linked Open Data - Creating Knowledge Out of Interlinked Data:
  Results of the LOD2 Project}}, volume 8661.
\newblock Springer, 2014.

\bibitem{austin2018path}
C.~C. Austin.
\newblock {A Path to Big Data Readiness}.
\newblock In {\em 2018 IEEE International Conference on Big Data (Big Data)},
  pages 4844--4853. IEEE, 2018.

\bibitem{Data61}
{Australia’s National Science Agency}.
\newblock {Data61: Australian "Digital Twin" Technology Set to Transform
  Manufacturing}.
\newblock
  \url{https://www.csiro.au/en/News/News-releases/2019/Australian-digital-twin-technology-set-to-transform-manufacturing}.
\newblock [Online; accessed 15-December-2022].

\bibitem{baheer2020systematic}
B.~A. Baheer, D.~Lamas, and S.~Sousa.
\newblock {A Systematic Literature Review on Existing Digital Government
  Architectures: State-of-the-Art, Challenges, and Prospects}.
\newblock {\em Administrative Sciences}, 10(2):25, 2020.

\bibitem{bakar2020digital}
K.~A. Bakar, R.~Ibrahim, and Y.~Yunus.
\newblock {Digital Government Evolution and Maturity Models: A Review}.
\newblock {\em Open International Journal of Informatics}, 8(2):70--87, 2020.

\bibitem{balashov2020prospects}
A.~Balashov, A.~Barabanov, V.~Degtereva, and M.~Ivanov.
\newblock {Prospects for Digital Transformation of Public Administration in
  Russia}.
\newblock In {\em Proceedings of the 2nd International Scientific Conference on
  Innovations in Digital Economy}, pages 1--7, 2020.

\bibitem{barcelona2017agile}
{Barcelona Ciutat Digital}.
\newblock {Agile Methodologies at Barcelona City Council}.
\newblock
  \url{https://www.barcelona.cat/digitalstandards/en/agile-methodologies/0.1/_attachments/barcelona_agile_methodologies_0.1.en.pdf}.
\newblock [Online; accessed 20-January-2023].

\bibitem{barns2016mine}
S.~Barns.
\newblock {Mine your Data: Open Data, Digital Strategies and Entrepreneurial
  Governance by Code}.
\newblock {\em Urban Geography}, 37(4):554--571, 2016.

\bibitem{barricelli2019survey}
B.~R. Barricelli, E.~Casiraghi, and D.~Fogli.
\newblock {A Survey on Digital Twin: Definitions, Characteristics,
  Applications, and Design Implications}.
\newblock {\em IEEE access}, 7:167653--167671, 2019.

\bibitem{DevOps}
L.~Bass, I.~Weber, and L.~Zhu.
\newblock {\em {DevOps: A Software Architect's Perspective}}.
\newblock Addison-Wesley Professional, 2015.

\bibitem{battisti2020digital}
D.~Battisti.
\newblock {The Digital Transformation of Italy’s Public Sector: Government
  Cannot Be Left Behind!}
\newblock {\em JeDEM-eJournal of eDemocracy and Open Government}, 12(1):25--39,
  2020.

\bibitem{besta2018survey}
M.~Besta and T.~Hoefler.
\newblock {Survey and Taxonomy of Lossless Graph Compression and
  Space-Efficient Graph Representations}.
\newblock {\em arXiv preprint arXiv:1806.01799}, 2018.

\bibitem{bogdanova2020agile}
M.~Bogdanova, E.~Parashkevova, and M.~Stoyanova.
\newblock {Agile Project Management in Public Sector - Methodological Aspects}.
\newblock {\em Journal of European Economy}, 19(2):283--298, 2020.

\bibitem{borg2018digitalization}
M.~Borg, T.~Olsson, U.~Franke, and S.~Assar.
\newblock {Digitalization of Swedish Government Agencies: A Perspective Through
  the Lens of a Software Development Census}.
\newblock In {\em Proceedings of the 40th International Conference on Software
  Engineering: Software Engineering in Society}, pages 37--46, 2018.

\bibitem{bounabatgovernment}
B.~Bounabat.
\newblock {From E-Government to Digital Government: Stakes and Evolution Models
  Du E-Gouvernement au Gouvernement Digital: Enjeux et Mod{\`e}les
  d’{\'E}volution}.
\newblock {\em Electronic Journal of Information Technology}, 10(1):1--20,
  2017.

\bibitem{bousdekis2020digital}
A.~Bousdekis and D.~Kardaras.
\newblock {Digital Transformation of Local Government: A Case Study from
  Greece}.
\newblock In {\em 2020 IEEE 22nd Conference on Business Informatics (CBI)},
  volume~2, pages 131--140. IEEE, 2020.

\bibitem{SPBportal}
{Brazilian Federal Government}.
\newblock {Brazilian Public Software: Operations Manual (prod)}.
\newblock \url{https://softwarepublico.gov.br/doc/}.
\newblock [Online; accessed 03-March-2023].

\bibitem{brisaboa2020space}
N.~R. Brisaboa, A.~Cerdeira-Pena, G.~de~Bernardo, A.~Fari{\~n}a, and
  G.~Navarro.
\newblock {Space/Time-Efficient RDF Stores Based on Circular Suffix Sorting}.
\newblock {\em arXiv preprint arXiv:2009.10045}, 2020.

\bibitem{bucena2017simplifying}
I.~Bucena and M.~Kirikova.
\newblock {Simplifying the DevOps Adoption Process}.
\newblock In {\em BIR Workshops}, pages 1--15, 2017.

\bibitem{calderon2018smartness}
M.~Calderón, G.~López, and G.~Marín.
\newblock {Smartness and Technical Readiness of Latin American Cities: A
  Critical Assessment}.
\newblock {\em IEEE Access}, 6:56839--56850, 2018.

\bibitem{charalabidis2018multiple}
Y.~Charalabidis, A.~Zuiderwijk, C.~Alexopoulos, M.~Janssen, T.~Lampoltshammer,
  and E.~Ferro.
\newblock {The Multiple Life Cycles of Open Data Creation and Use}.
\newblock In {\em The World of Open Data}, pages 11--31. Springer, 2018.

\bibitem{chaudhuryreforming}
K.~Chaudhury, A.~Barua, R.~Deka, T.~Gogoi, A.~C. Gupta, and S.~Pyarelal.
\newblock {Reforming and Strengthening Digital Service Delivery: Case of
  Government of Assam}, 2020.

\bibitem{Chicago_Taxi_TNP}
{Chicago Open Data Portal Team}.
\newblock {How Chicago Protects Privacy in TNP and TAXI Open Data}.
\newblock
  \url{http://dev.cityofchicago.org/open\%20data/data\%20portal/2019/04/12/tnp-taxi-privacy.html}.
\newblock [Online; accessed 02-September-2022].

\bibitem{Chicago_OGD_Data}
{City of Chicago}.
\newblock {Chicago Open Data Portal}.
\newblock \url{http://data.cityofchicago.org/}.
\newblock [Online; accessed 02-September-2022].

\bibitem{Chicago_OGD_Team}
{City of Chicago}.
\newblock {Chicago Open Data Portal Development Team}.
\newblock \url{https://dev.cityofchicago.org/}.
\newblock [Online; accessed 02-September-2022].

\bibitem{Chicago_FI_ML}
{City of Chicago}.
\newblock {Food Inspections Evaluation Machine Learning Model}.
\newblock \url{https://github.com/Chicago/food-inspections-evaluation}.
\newblock [Online; accessed 12-September-2022].

\bibitem{Chicago_foodborne_illness}
{City of Chicago}.
\newblock {Reduction the Exposure of Citizens of Chicago to Foodborne
  Diseases}.
\newblock
  \url{https://www.chicago.gov/city/en/sites/gearupgetready/home/foodborne-illness.html}.
\newblock [Online; accessed 12-September-2022].

\bibitem{reis2021governance}
L.~Claudio Diogo~Reis, F.~Cristina~Bernardini, S.~Bacellar Leal~Ferreira, and
  C.~Cappelli.
\newblock {ICT Governance in Brazilian Smart Cities: An Integrative Approach in
  the Context of Digital Transformation}.
\newblock In {\em DG.O2021: The 22nd Annual International Conference on Digital
  Government Research}, DG.O'21, page 302–316, New York, NY, USA, 2021. ACM.

\bibitem{clusit2022report}
CLUSIT.
\newblock {Italy Report on ICT Security}.
\newblock Technical report, Italian Association for IT Security, 2022.

\bibitem{CMMI}
{CMMI Institute}.
\newblock {Capability Digital Maturity Model (CMMI V2.0)}.
\newblock \url{https://cmmiinstitute.com/cmmi}.
\newblock [Online; accessed 30-September-2022].

\bibitem{CNIL_map}
{Commission Nationale de l'Informatique et des Libertés (CNIL)}.
\newblock {CNIL: Map of the Data Protection Authorities Around the World}.
\newblock \url{https://www.cnil.fr/en/data-protection-around-the-world}.
\newblock [Online; accessed 17-October-2022].

\bibitem{cordella2019government}
A.~Cordella and A.~Paletti.
\newblock {Government as a Platform, Orchestration, and Public Value Creation:
  The Italian Case}.
\newblock {\em Government Information Quarterly}, 36(4):101409, 2019.

\bibitem{GI_Index}
{Cornell University, Institut Européen d’Administration des Affairs, and
  World Intellectual Property Organization}.
\newblock {Global Innovation Index}.
\newblock \url{https://www.globalinnovationindex.org/}.
\newblock [Online; accessed 30-September-2022].

\bibitem{dai2016data}
W.~Dai, I.~Wardlaw, Y.~Cui, K.~Mehdi, Y.~Li, and J.~Long.
\newblock {Data Profiling Technology of Data Governance Regarding Big Data:
  Review and Rethinking}.
\newblock {\em Information Technology: New Generations}, pages 439--450, 2016.

\bibitem{danielsen2021benefits}
F.~Danielsen.
\newblock {Benefits and Challenges of Digitalization: An Expert Study on
  Norwegian Public Organizations}.
\newblock In {\em DG.O2021: The 22nd Annual International Conference on Digital
  Government Research}, pages 317--326, 2021.

\bibitem{datta2020digital}
P.~Datta, L.~Walker, and F.~Amarilli.
\newblock {Digital Transformation: Learning from Italy’s Public
  Administration}.
\newblock {\em Journal of Information Technology Teaching Cases}, 10(2):54--71,
  2020.

\bibitem{davoudian2020big}
A.~Davoudian and M.~Liu.
\newblock {Big Data Systems: A Software Engineering Perspective}.
\newblock {\em ACM Computing Surveys (CSUR)}, 53(5):1--39, 2020.

\bibitem{de2019evaluating}
V.~De~Andrade~Soares, G.~Yukari~Iwama, V.~Gomes~Menezes, M.~Miranda
  Forte~Gomes, G.~Vitor~Pedrosa, W.~CM~Pereira~da Silva, and R.~Maria Da
  Costa~Figueiredo.
\newblock {Evaluating Government Services Based on User Perspective}.
\newblock In {\em Proceedings of the 20th Annual International Conference on
  Digital Government Research}, pages 425--432, 2019.

\bibitem{barcelona2018procurement}
{\`A}.~de~Treball, F.~Bria, P.~Rodriguez, M.~Bain, J.~Batlle, A.~Bastida~Vila,
  X.~E. Barandiaran~Fern{\'a}ndez, M.~Boada~Pla, G.~Marpons~Ucero,
  X.~Roca~Vilalta, et~al.
\newblock {Barcelona City Council ICT Public Procurement Guide}, 2018.

\bibitem{dener2021govtech}
C.~Dener, H.~Nii-Aponsah, L.~E. Ghunney, and K.~D. Johns.
\newblock {\em {GovTech Maturity Index: The State of Public Sector Digital
  Transformation}}.
\newblock World Bank Publications, 2021.

\bibitem{DWZ2021}
L.~Deren, Y.~Wenbo, and S.~Zhenfeng.
\newblock {Smart City Based on Digital Twins}.
\newblock {\em Computational Urban Science}, 1(4):11, 2021.

\bibitem{CAF_Index}
{Development Bank of Latin America}.
\newblock {Development Bank of Latin America GovTech Index}.
\newblock
  \url{https://www.caf.com/en/currently/news/2020/06/caf-publishes-the-first-iberoamerican-govtech-index/}.
\newblock [Online; accessed 30-September-2022].

\bibitem{dwivedula2020transitioning}
R.~Dwivedula and N.~Bolloju.
\newblock {Transitioning from Plan-Driven Methods to Agile Methods-Preparation
  for a Systematic Literature Review}.
\newblock In {\em 2020 5th International Conference on Communication and
  Electronics Systems (ICCES)}, pages 944--950. IEEE, 2020.

\bibitem{eaton2015distributed}
B.~Eaton, S.~Elaluf-Calderwood, C.~S{\o}rensen, and Y.~Yoo.
\newblock {Distributed Tuning of Boundary Resources}.
\newblock {\em MIS Quarterly}, 39(1):217--244, 2015.

\bibitem{edison2021comparing}
H.~Edison, X.~Wang, and K.~Conboy.
\newblock {Comparing Methods for Large-Scale Agile Software Development: A
  systematic Literature Review}.
\newblock {\em IEEE Transactions on Software Engineering}, 2021.

\bibitem{dcat-ap}
{European Commission}.
\newblock {DCAT Application Profile for Data Portals in Europe}.
\newblock
  \url{https://joinup.ec.europa.eu/collection/semantic-interoperability-community-semic/solution/dcat-application-profile-data-portals-europe}.
\newblock [Online; accessed 21-February-2023].

\bibitem{DigEducActPlan2027}
{European Commission}.
\newblock {Digital Education Action Plan (2021-2027)}.
\newblock
  \url{https://education.ec.europa.eu/focus-topics/digital-education/action-plan}.
\newblock [Online; accessed 20-March-2023].

\bibitem{EC_eGov_Benchmark_Index}
{European Commission}.
\newblock {EC eGovernment Benchmark}.
\newblock
  \url{https://digital-strategy.ec.europa.eu/en/library/egovernment-benchmark-2022}.
\newblock [Online; accessed 30-September-2022].

\bibitem{STORK_project}
{European Commission}.
\newblock {Secure idenTity acrOss boRders linKed (STORK) European Project}.
\newblock
  \url{https://joinup.ec.europa.eu/collection/secure-identity-across-borders-linked-stork}.
\newblock [Online; accessed 10-October-2022].

\bibitem{eucomm2016}
{European Commission}.
\newblock {European Interoperability Framework - Implementation Strategy
  (2017-2020)}.
\newblock Luxembourg: Publications Office of the European Union, 2016.

\bibitem{GDPR}
{European Commission}.
\newblock {General Data Protection Regulation (GDPR)}.
\newblock \url{https://eur-lex.europa.eu/eli/reg/2016/679}, 2022.
\newblock [Online; accessed 25-June-2022].

\bibitem{ENISA}
{European Union Government}.
\newblock {European Union Agency for Cybersecurity (ENISA) Website}.
\newblock \url{https://www.enisa.europa.eu/}.
\newblock [Online; accessed 08-October-2022].

\bibitem{fagarasan2022delivery}
C.~Fagarasan, C.~Cristea, M.~Cristea, O.~Popa, C.~Mihele, and A.~Pisla.
\newblock {The Delivery of Large-Scale Software Products Through the Adoption
  of the SAFe Framework}.
\newblock In {\em 2022 International Conference on Development and Application
  Systems (DAS)}, pages 137--143. IEEE, 2022.

\bibitem{fangmann2020agile}
J.~Fangmann, H.~Looks, J.~Thomaschewski, and E.-M. Sch{\"o}n.
\newblock {Agile Transformation in E-Government Projects}.
\newblock In {\em 2020 15th Iberian Conference on Information Systems and
  Technologies (CISTI)}, pages 1--4. IEEE, 2020.

\bibitem{farshchian2020experiences}
B.~A. Farshchian, H.~S{\ae}tertr{\o}, and M.~St{\aa}laker.
\newblock {Experiences From an Interpretative Case Study of Innovative Public
  Procurement of Digital Systems in the Norwegian Public Sector}.
\newblock In {\em Procs. the Int. Conf. on Evaluation and Assessment in
  Software Engineering EASE}, ACM International Conference Proceeding Series,
  pages 373--374, 2020.

\bibitem{feher2018digitalization}
P.~Feh{\'e}r and Z.~Szab{\'o}.
\newblock {Digitalization in the Public Sector - Findings of a Hungarian
  Survey}.
\newblock In {\em 2018 12th International Conference on Software, Knowledge,
  Information Management \& Applications (SKIMA)}, pages 1--6. IEEE, 2018.

\bibitem{filograna2016cloudification}
A.~Filograna, P.~Smiraglia, C.~Gilsanz, S.~Krco, A.~Medela, and T.~Su.
\newblock {Cloudification of Public Services in Smart Cities: The Clips
  Project}.
\newblock In {\em Proc. IEEE Symposium on Computers and Communication (ISCC)},
  pages 153--158. IEEE, 2016.

\bibitem{forsgren2018devops}
N.~Forsgren and M.~Kersten.
\newblock {DevOps Metrics}.
\newblock {\em Communications of the ACM}, 61(4):44--48, 2018.

\bibitem{fuller2020digital}
A.~Fuller, Z.~Fan, C.~Day, and C.~Barlow.
\newblock {Digital Twin: Enabling Technologies, Challenges and Open Research}.
\newblock {\em IEEE access}, 8:108952--108971, 2020.

\bibitem{ghimire2020scaling}
D.~Ghimire, S.~Charters, and S.~Gibbs.
\newblock {Scaling Agile Software Development Approach in Government
  Organization in New Zealand}.
\newblock In {\em Proceedings of the 3rd International Conference on Software
  Engineering and Information Management}, pages 100--104, 2020.

\bibitem{gil2015makes}
J.~R. Gil-Garcia, T.~A. Pardo, and T.~Nam.
\newblock {What Makes a City Smart? Identifying Core Components and Proposing
  an Integrative and Comprehensive Conceptualization}.
\newblock {\em Information Polity}, 20(1):61--87, 2015.

\bibitem{gong2020towards}
Y.~Gong, J.~Yang, and X.~Shi.
\newblock {Towards a Comprehensive Understanding of Digital Transformation in
  Government: Analysis of Flexibility and Enterprise Architecture}.
\newblock {\em Government Information Quarterly}, 37(3):101487, 2020.

\bibitem{guerreri2021evolving}
P.~M. Guerrieri, S.~Comai, and M.~Fugini.
\newblock {Evolving Experiences of Participation: Towards an e-ILAUD Tool}.
\newblock In {\em 2021 IEEE International Smart Cities Conference (ISC2)},
  pages 1--7, 2021.

\bibitem{habibzadeh2019smart}
H.~Habibzadeh, C.~Kaptan, T.~Soyata, B.~Kantarci, and A.~Boukerche.
\newblock {Smart City System Design: A Comprehensive Study of the Application
  and Data Planes}.
\newblock {\em ACM Computing Surveys}, 52(2), may 2019.

\bibitem{harrison2019applying}
T.~Harrison, D.~Canestraro, T.~Pardo, M.~Avila-Maravilla, N.~Soto,
  M.~Sutherland, G.~B. Burke, and M.~Gasco.
\newblock {Applying an Enterprise Data Model in Government}.
\newblock In {\em Proceedings of the 20th Annual International Conference on
  Digital Government Research}, pages 265--271, 2019.

\bibitem{hastings2019unlocking}
J.~S. Hastings, M.~Howison, T.~Lawless, J.~Ucles, and P.~White.
\newblock {Unlocking Data to Improve Public Policy}.
\newblock {\em Communications of the ACM}, 62(10):48--53, 2019.

\bibitem{hawken2020open}
S.~Hawken, H.~Han, and C.~Pettit.
\newblock {\em {Open Cities | Open Data: Collaborative Cities in the
  Information Era}}.
\newblock Springer, 2020.

\bibitem{hedestig2018co}
U.~Hedestig, D.~Skog, and M.~S{\"o}derstr{\"o}m.
\newblock {Co-Producing Public Value Through IoT and Social Media}.
\newblock In {\em Proceedings of the 19th Annual International Conference on
  Digital Government Research: Governance in the Data Age}, pages 1--10, 2018.

\bibitem{hermanto2020improving}
A.~Hermanto, R.~B. Ibrahim, and G.~Kusnanto.
\newblock {Improving Value-Based E-Government Towards the Achievement of Smart
  Government}.
\newblock In {\em 2020 Fifth International Conference on Informatics and
  Computing (ICIC)}, pages 1--7. IEEE, 2020.

\bibitem{hsiao2019elevated}
Y.-C. Hsiao, M.-H. Wu, and S.~C. Li.
\newblock {Elevated Performance of the Smart City — A Case Study of the IoT
  by Innovation Mode}.
\newblock {\em IEEE Transactions on Engineering Management}, 68(5):1461--1475,
  2019.

\bibitem{i2017government}
{i Hisenda, Comissi{\'o} d'Economia and de Treball, {\`A}rea and i Economia,
  Ger{\`e}ncia de Presid{\`e}ncia and others}.
\newblock {A Government Measure for Open Digitisation: Free Software and Agile
  Development of Public Administration Services}.
\newblock {\em Barcelona Ciutat Digital}, 2017.

\bibitem{idowu2021framework}
L.~Idowu~Lamid, I.~Ali~Ibrahim, K.~Inuwa~Abdullahi, and U.~Gambo~Abdullahi.
\newblock {A Framework for Digital Government Transformation Performance
  Assessment and Toolkit for Developing Countries}.
\newblock In {\em Proceedings of the 14th International Conference on Theory
  and Practice of Electronic Governance}, pages 203--215, 2021.

\bibitem{isin2020being}
E.~Isin and E.~Ruppert.
\newblock {\em {Being Digital Citizens}}.
\newblock Rowman \& Littlefield Publishers, 2 edition, 2020.

\bibitem{ISOIEC_jtc1}
{ISO/IEC JTC1 Information Technology}.
\newblock {Smart Cities - Preliminary Report 2014}.
\newblock {ISO/IEC}, 2015.

\bibitem{Nat_CS_Agency_IT}
{Italian Government}.
\newblock {National Cybersecurity Agency (ACN)}.
\newblock \url{https://www.acn.gov.it/en}.
\newblock [Online; accessed 05-September-2022].

\bibitem{NRRP_IT}
{Italian Ministry of Economy and Finance (MEF)}.
\newblock {The National Recovery and Resilience Plan (NRRP)}.
\newblock
  \url{https://www.mef.gov.it/en/focus/The-National-Recovery-and-Resilience-Plan-NRRP/}.
\newblock [Online; accessed 27-September-2022].

\bibitem{jadi2017implementation}
Y.~Jadi and L.~Jie.
\newblock {An Implementation Framework of Business Intelligence in E-Government
  Systems for Developing countries: Case study: Morocco E-Government System}.
\newblock In {\em 2017 International Conference on Information Society
  (i-Society)}, pages 138--142. IEEE, 2017.

\bibitem{janssen2012benefits}
M.~Janssen, Y.~Charalabidis, and A.~Zuiderwijk.
\newblock {Benefits, Adoption Barriers and Myths of Open Data and Open
  Government}.
\newblock {\em Information Systems Management}, 29(4):258--268, 2012.

\bibitem{javed2020biotope}
A.~Javed, S.~Kubler, A.~Malhi, A.~Nurminen, J.~Robert, and K.~Fr{\"a}mling.
\newblock Biotope: Building an iot open innovation ecosystem for smart cities.
\newblock {\em IEEE Access}, 8:224318--224342, 2020.

\bibitem{jonathan2019digital}
G.~M. Jonathan.
\newblock Digital transformation in the public sector: Identifying critical
  success factors.
\newblock In {\em European, Mediterranean, and Middle Eastern Conference on
  Information Systems}, pages 223--235. Springer, 2019.

\bibitem{jonathan2020privacy}
G.~M. Jonathan, B.~K. Gebremeskel, and S.~D. Yalew.
\newblock {Privacy and Security in the Digitalisation Era}.
\newblock In {\em 2020 11th IEEE Annual Information Technology, Electronics and
  Mobile Communication Conference (IEMCON)}, pages 0837--0844. IEEE, 2020.

\bibitem{jovanovic2020agile}
M.~Jovanovi{\'c}, A.-L. Mesquida, A.~Mas, and R.~Colomo-Palacios.
\newblock {Agile Transition and Adoption Frameworks, Issues and Factors: A
  Systematic Mapping}.
\newblock {\em IEEE Access}, 8:15711--15735, 2020.

\bibitem{JLA22}
G.~Juell-Skielse, I.~Lindgren, and M.~Akesson.
\newblock {\em {Service Automation in the Public Sector - Concepts, Empirical
  Examples and Challenges}}.
\newblock Springer, 2022.

\bibitem{juniawan2017smart}
M.~A. Juniawan, P.~Sandhyaduhita, B.~Purwandari, S.~B. Yudhoatmojo, and
  M.~A.~A. Dewi.
\newblock {Smart Government Assessment Using Scottish Smart City Maturity
  Model: A Case Study of Depok City}.
\newblock In {\em 2017 International Conference on Advanced Computer Science
  and Information Systems (ICACSIS)}, pages 99--104. IEEE, 2017.

\bibitem{jussupova2019digital}
G.~Jussupova, B.~Bokayev, and D.~Zhussip.
\newblock {Digital Government Maturity as a Technologically New E-Government
  Maturity Model: Experience of Kazakhstan}.
\newblock In {\em Proceedings of the 2019 3rd International Conference on
  E-commerce, E-Business and E-Government}, pages 10--14, 2019.

\bibitem{kalogirou2020linked}
V.~Kalogirou, S.~Van~Dooren, I.~Dimopoulos, Y.~Charalabidis, J.-P. De-Baets,
  and G.~Lobo.
\newblock {Linked Government Data Hub, an Ontology Agnostic Data Harvester and
  API}.
\newblock In {\em Proceedings of the 13th International Conference on Theory
  and Practice of Electronic Governance}, pages 779--782, 2020.

\bibitem{kalvet2018contributing}
T.~Kalvet, M.~Toots, and R.~Krimmer.
\newblock {Contributing to a Digital Single Market for Europe: Barriers and
  Drivers of an EU-Wide Once-Only Principle}.
\newblock In {\em Proceedings of the 19th Annual International Conference on
  Digital Government Research: Governance in the Data Age}, pages 1--8, 2018.

\bibitem{kane2015strategy}
G.~C. Kane, D.~Palmer, A.~N. Phillips, D.~Kiron, N.~Buckley, et~al.
\newblock {Strategy, Not Technology, Drives Digital Transformation}.
\newblock {\em MIT Sloan Management Review and Deloitte University Press},
  14(1-25), 2015.

\bibitem{Open_Data}
R.~Kitchin.
\newblock {\em {The Data Revolution: Big Data, Open Data, Data Infrastructures
  and Their Consequences}}.
\newblock Sage, 2014.

\bibitem{kitsing2018janus}
M.~Kitsing.
\newblock The janus-faced approach to governance: a mismatch between public
  sector reforms and digital government in estonia.
\newblock In {\em Proceedings of the 11th International Conference on Theory
  and Practice of Electronic Governance}, pages 59--68, 2018.

\bibitem{kitsing2019alternative}
M.~Kitsing.
\newblock {Alternative Futures for Digital Governance}.
\newblock In {\em Proceedings of the 20th Annual International Conference on
  Digital Government Research}, pages 48--59, 2019.

\bibitem{kuvcera2013open}
J.~Ku{\v{c}}era, D.~Chlapek, and M.~Ne{\v{c}}ask{\`y}.
\newblock {Open Government Data Catalogs: Current Approaches and Quality
  Perspective}.
\newblock In {\em International Conference on Electronic Government and the
  Information Systems Perspective}, pages 152--166. Springer, 2013.

\bibitem{kupi2021late}
M.~Kupi.
\newblock {Late to the Party? Agile Methods in British and German Government
  Institutions}.
\newblock {\em SocArXiv}, 2021.

\bibitem{lammel2020metadata}
P.~L{\"a}mmel, B.~Dittwald, L.~Bruns, N.~Tcholtchev, Y.~Glikman, S.~Cuno,
  M.~Fl{\"u}gge, and I.~Schieferdecker.
\newblock {Metadata Harvesting and Quality Assurance Within Open Urban
  Platforms}.
\newblock {\em Journal of Data and Information Quality (JDIQ)}, 12(4):1--20,
  2020.

\bibitem{lanza2020what}
B.~B.~B. Lanza, J.~R. Gil-Garcia, and T.~A. Pardo.
\newblock {What Makes a City Smart? Reconsidering the Core Components in the
  Brazilian Context}.
\newblock In {\em Proceedings of the 13th International Conference on Theory
  and Practice of Electronic Governance}, ICEGOV '20, page 638–645, New York,
  NY, USA, 2020. ACM.

\bibitem{leao2018digitization}
H.~A.~T. Le{\~a}o and E.~D. Canedo.
\newblock {Digitization of Public Services: A Systematic Literature Review}.
\newblock In {\em Proceedings of the 17th Brazilian Symposium on Software
  Quality}, pages 91--100, 2018.

\bibitem{legner2017digitalization}
C.~Legner, T.~Eymann, T.~Hess, C.~Matt, T.~B{\"o}hmann, P.~Drews,
  A.~M{\"a}dche, N.~Urbach, and F.~Ahlemann.
\newblock {Digitalization: Opportunity and Challenge for the Business and
  Information Systems Engineering Community}.
\newblock {\em Business \& Information Systems Engineering}, 59(4):301--308,
  2017.

\bibitem{lepekhin2019systematic}
A.~Lepekhin, A.~Borremans, I.~Ilin, and S.~Jantunen.
\newblock {A Systematic Mapping Study on Internet of Things Challenges}.
\newblock In {\em 2019 IEEE/ACM 1st International Workshop on Software
  Engineering Research \& Practices for the Internet of Things (SERP4IoT)},
  pages 9--16. IEEE, 2019.

\bibitem{li2018keys}
H.-L. Li.
\newblock {Keys to Promoting Governmental Digital Transformation}.
\newblock {\em International Journal of Automation and Smart Technology},
  8(4):145--148, 2018.

\bibitem{lindgren2018digital}
I.~Lindgren and A.~F. van Veenstra.
\newblock {Digital Government Transformation: A Case Illustrating Public
  E-Service Development as Part of Public Sector Transformation}.
\newblock In {\em Proceedings of the 19th Annual International Conference on
  Digital Government Research: Governance in the Data Age}, pages 1--6, 2018.

\bibitem{liu2022citizen}
C.~Liu and D.~Zowghi.
\newblock {Citizen Involvement in Digital Transformation: A Systematic Review
  and a Framework}.
\newblock {\em Online Information Review}, ahead-of-print(ahead-of-print),
  2022.

\bibitem{liu2018graph}
Y.~Liu, T.~Safavi, A.~Dighe, and D.~Koutra.
\newblock {Graph Summarization Methods and Applications: A Survey}.
\newblock {\em ACM computing surveys (CSUR)}, 51(3):1--34, 2018.

\bibitem{lopes2018business}
F.~J.~R. Lopes and M.~I. Cort{\'e}s.
\newblock {Business Value Characterization in Software Projects for Electronic
  Government in the Brazilian Federal Government}.
\newblock In {\em Proceedings of the XXXII Brazilian Symposium on Software
  Engineering}, pages 82--91, 2018.

\bibitem{lopes2018public}
N.~V. Lopes and S.~B. Dhaou.
\newblock {Public Service Delivery Framework: Case of Canada, China and
  Estonia}.
\newblock In {\em Proc. 11th International Conference on Theory and Practice of
  Electronic Governance}, pages 101--110, 2018.

\bibitem{luciano2020role}
E.~M. Luciano and G.~C. Wiedenh\"{o}ft.
\newblock {The Role of Organizational Citizenship Behavior and Strategic
  Alignment in Increasing the Generation of Public Value Through Digital
  Transformation}.
\newblock In {\em Proceedings of the 13th International Conference on Theory
  and Practice of Electronic Governance}, ICEGOV '20, page 494–501, New York,
  NY, USA, 2020. ACM.

\bibitem{luckner2020urban}
M.~Luckner, M.~Grzenda, R.~Kunicki, and J.~Legierski.
\newblock {IoT Architecture for Urban Data-Centric Services and Applications}.
\newblock {\em ACM Transactions on Internet Technology}, 20(3), 2020.

\bibitem{luque2020urban}
A.~Luque-Ayala and S.~Marvin.
\newblock {\em {Urban Operating Systems: Producing the Computational City}}.
\newblock MIT Press, 2020.

\bibitem{luz2018experience}
W.~Luz, E.~Agilar, M.~C. de~Oliveira, C.~E.~R. de~Melo, G.~Pinto, and
  R.~Bonif{\'a}cio.
\newblock {An Experience Report on the Adoption of Microservices in Three
  Brazilian Government Institutions}.
\newblock In {\em Proceedings of the XXXII Brazilian Symposium on Software
  Engineering}, pages 32--41, 2018.

\bibitem{luz2018building}
W.~P. Luz, G.~Pinto, and R.~Bonif{\'a}cio.
\newblock {Building a Collaborative Culture: A Grounded Theory of Well
  Succeeded DevOps Adoption in Practice}.
\newblock In {\em Proc. 12th ACM/IEEE Int. Symposium on Empirical Software
  Engineering and Measurement}, pages 1--10, 2018.

\bibitem{luz2019adopting}
W.~P. Luz, G.~Pinto, and R.~Bonif{\'a}cio.
\newblock {Adopting DevOps in the Real World: A Theory, a Model, and a Case
  Study}.
\newblock {\em Journal of Systems and Software}, 157:110384, 2019.

\bibitem{mahraz2019systematic}
M.-I. Mahraz, L.~Benabbou, and A.~Berrado.
\newblock {A Systematic Literature Review of Digital Transformation}.
\newblock In {\em International Conference on Industrial Engineering and
  Operations Management. IEOM Society International}, pages 917--931, 2019.

\bibitem{malhotra2019designing}
C.~Malhotra, V.~Kotwal, and A.~Basu.
\newblock {Designing National Health Stack for Public Health: Role of ICT-Based
  Knowledge Management System}.
\newblock In {\em 2019 ITU Kaleidoscope: ICT for Health: Networks, Standards
  and Innovation (ITU K)}, pages 1--8. IEEE, 2019.

\bibitem{manda2021leadership}
M.~I. Manda.
\newblock {Leadership and Trust as Key Pillars in “Smart Governance” for
  Inclusive Growth in the 4th Industrial Revolution (4IR): Evidence from South
  Africa}.
\newblock In {\em Proceedings of the 14th International Conference on Theory
  and Practice of Electronic Governance}, pages 308--315, 2021.

\bibitem{maneth2015survey}
S.~Maneth and F.~Peternek.
\newblock {A Survey on Methods and Systems for Graph Compression}.
\newblock {\em arXiv preprint arXiv:1504.00616}, 2015.

\bibitem{mantovani2020characteristics}
R.~Mantovani~Fontana and S.~Marczak.
\newblock {Characteristics and Challenges of Agile Software Development
  Adoption in Brazilian Government}.
\newblock {\em Journal of Technology Management \& Innovation}, 15(2):3--10,
  2020.

\bibitem{margariti2020assessment}
V.~Margariti, D.~Anagnostopoulos, A.~Papastilianou, T.~Stamati, and S.~Angeli.
\newblock {Assessment of Organizational Interoperability in E-Government: A New
  Model and Tool for Assessing Organizational Interoperability Maturity of a
  Public Service in Practice}.
\newblock In {\em Proceedings of the 13th International Conference on Theory
  and Practice of Electronic Governance}, pages 298--308, 2020.

\bibitem{margetts2013second}
H.~Margetts and P.~Dunleavy.
\newblock {The Second Wave of Digital-Era Governance: A Quasi-Paradigm for
  Government on the Web}.
\newblock {\em Philosophical Transactions of the Royal Society A: Mathematical,
  Physical and Engineering Sciences}, 371(1987):20120382, 2013.

\bibitem{maroukian2020leading}
K.~Maroukian and S.~R. Gulliver.
\newblock {Leading DevOps Practice and Principle Adoption}.
\newblock {\em arXiv preprint arXiv:2008.10515}, 2020.

\bibitem{MartPrieto2018}
M.~A. Mart{\'i}nez-Prieto, J.~D. Fern{\'a}ndez, A.~Hern{\'a}ndez-Illera, and
  C.~Guti{\'e}rrez.
\newblock {\em {RDF Compression}}, pages 1--11.
\newblock Springer International Publishing, 2018.

\bibitem{masombuka2020framework}
K.~T. Masombuka.
\newblock {\em {A Framework for a Successful Collaboration Culture in Software
  Development and Operations (DevOps) Environments}}.
\newblock PhD thesis, School of Computing, University of South Africa, 2020.

\bibitem{mcbride2019does}
K.~McBride, G.~Aavik, M.~Toots, T.~Kalvet, and R.~Krimmer.
\newblock {How Does Open Government Data Driven Co-Creation Occur? Six Factors
  and a ‘Perfect Storm’; Insights From Chicago's Food Inspection
  Forecasting Model}.
\newblock {\em Government Information Quarterly}, 36(1):88--97, 2019.

\bibitem{meirelles2017brazilian}
P.~Meirelles, M.~Wen, A.~Terceiro, R.~Siqueira, L.~Kanashiro, and H.~Neri.
\newblock {Brazilian Public Software Portal: An Integrated Platform for
  Collaborative Development}.
\newblock In {\em Proceedings of the 13th International Symposium on Open
  Collaboration}, pages 1--10, 2017.

\bibitem{menezes2022evaluation}
V.~G.~d. Menezes, G.~V. Pedrosa, M.~P.~d. Silva, and R.~M. d.~C. Figueiredo.
\newblock {Evaluation of Public Services Considering the Expectations of Users
  — A Systematic Literature Review}.
\newblock {\em Information}, 13(4):162, 2022.

\bibitem{mergel2021agile}
I.~Mergel, S.~Ganapati, and A.~B. Whitford.
\newblock {Agile: A New Way of Governing}.
\newblock {\em Public Administration Review}, 81(1):161--165, 2021.

\bibitem{mergel2018citizen}
I.~Mergel, R.~Kattel, V.~Lember, and K.~McBride.
\newblock {Citizen-Oriented Digital Transformation in the Public Sector}.
\newblock In {\em Proceedings of the 19th Annual International Conference on
  Digital Government Research: Governance in the Data Age}, pages 1--3, 2018.

\bibitem{meyerhoff2020digital}
M.~Meyerhoff~Nielsen and Z.~Jordanoski.
\newblock {Digital Transformation, Governance and Coordination Models: A
  Comparative Study of Australia, Denmark and the Republic of Korea}.
\newblock In {\em The 21st Annual International Conference on Digital
  Government Research}, pages 285--293, 2020.

\bibitem{miranda2021behind}
L.~F. Miranda~Ramos, M.~Lameiras, D.~Soares, and L.~Amaral.
\newblock {Who Is Behind the Scenes of the ICT Backstage? A Study of the ICT
  Resources in Local Governments}.
\newblock In {\em Proceedings of the 14th International Conference on Theory
  and Practice of Electronic Governance}, pages 165--171, 2021.

\bibitem{mohagheghi2021organizational}
P.~Mohagheghi and C.~Lassenius.
\newblock {Organizational Implications of Agile Adoption: A Case Study From the
  Public Sector}.
\newblock In {\em Proceedings of the 29th ACM Joint Meeting on European
  Software Engineering Conference and Symposium on the Foundations of Software
  Engineering}, pages 1444--1454, 2021.

\bibitem{monge2022new}
F.~Monge, B.-H. C.~L. Initiative, S.~Barns, R.~Kattel, and F.~Bria.
\newblock {A New Data Deal: The Case of Barcelona}.
\newblock Technical report, UCL Institute for Innovation and Public Purpose,
  Working Paper Series (No. WP 2022/02), 2022.

\bibitem{moustaka2018systematic}
V.~Moustaka, A.~Vakali, and L.~G. Anthopoulos.
\newblock {A Systematic Review for Smart City Data Analytics}.
\newblock {\em ACM Computing Surveys (cSuR)}, 51(5):1--41, 2018.

\bibitem{mubarkoot2021assessment}
M.~Mubarkoot.
\newblock {Assessment of Factors Influencing Adoption of DevOps Practices in
  Public Sector and Their Impact on Organizational Culture}.
\newblock In {\em Proceeding International Conference on Science and Technology
  (ICST)}, volume~2, pages 475--483, 2021.

\bibitem{mydyti2020cloud}
H.~Mydyti, J.~Ajdari, and X.~Zenuni.
\newblock {Cloud-Based Services Approach as Accelerator in Empowering Digital
  Transformation}.
\newblock In {\em 2020 43rd International Convention on Information,
  Communication and Electronic Technology (MIPRO)}, pages 1390--1396. IEEE,
  2020.

\bibitem{nachit2021digital}
H.~Nachit, M.~Jaafari, I.~El~Fikri, and L.~Belhcen.
\newblock {Digital Transformation in the Moroccan Public Sector: Drivers and
  Barriers}.
\newblock {\em SSRN}, 3907290, 2021.

\bibitem{nerima2021towards}
M.~Nerima and J.~Ralyt{\'e}.
\newblock {Towards a Digital Maturity Balance Model for Public Organizations}.
\newblock In {\em International Conference on Research Challenges in
  Information Science}, pages 295--310. Springer, 2021.

\bibitem{nerurkar2017agile}
A.~Nerurkar and I.~Das.
\newblock {Agile Project Management in Large Scale Digital Transformation
  Projects in Government and Public Sector: A Case Study of DILRMP Project}.
\newblock In {\em Proceeding 10th International Conference on Theory and
  Practice of Electronic Governance}, pages 580--581. ACM, 2017.

\bibitem{nerurkar2017analysis}
A.~Nerurkar and I.~Das.
\newblock {Analysis of DILRMP Project: Identifying the Applicability of Agile
  Project Management for Digital Transformation Projects in Government and
  Public Sector}.
\newblock In {\em Procs. the Special Collection on eGovernment Innovations in
  India}, pages 34--38. ACM, 2017.

\bibitem{neumann2022exploring}
O.~Neumann, K.~Guirguis, and R.~Steiner.
\newblock {Exploring Artificial Intelligence Adoption in Public Organizations:
  A Comparative Case Study}.
\newblock {\em Public Management Review}, pages 1--27, 2022.

\bibitem{NSW_DT}
{New South Wales Government (NSW)}.
\newblock {NSW Digital Twin}.
\newblock
  \url{https://www.digital.nsw.gov.au/article/twinning-spatial-services-has-created-a-digital-twin-of-nsw}.
\newblock [Online; accessed 15-December-2022].

\bibitem{nie2018chief}
Y.~Nie, J.~Talburt, X.~Li, and Z.~Xiao.
\newblock {Chief Data Officer (CDO) Role and Responsibility Analysis}.
\newblock {\em Journal of Computing Sciences in Colleges}, 33(5):4--12, 2018.

\bibitem{nikiforova2021open}
A.~Nikiforova and K.~McBride.
\newblock {Open Government Data Portal Usability: A User-Centred Usability
  Analysis of 41 Open Government Data Portals}.
\newblock {\em Telematics and Informatics}, 58:101539, 2021.

\bibitem{OECD}
{Organisation for Economic Co-Operation and Development (OECD)}.
\newblock \url{https://www.oecd.org/}.
\newblock [Online; accessed 30-September-2022].

\bibitem{OECD_Index}
{Organisation for Economic Co-Operation and Development (OECD)}.
\newblock {OECD Digital Government Index}.
\newblock \url{https://www.oecd.org/gov/digital-government/}.
\newblock [Online; accessed 30-September-2022].

\bibitem{ortiz2021design}
J.~Ortiz-Bejar and J.~Ortiz-Bejar.
\newblock {Design and Implementation of Digital Platform for E-Government}.
\newblock In {\em 2021 IEEE URUCON}, pages 547--551. IEEE, 2021.

\bibitem{osagie2017usability}
E.~Osagie, M.~Waqar, S.~Adebayo, A.~Stasiewicz, L.~Porwol, and A.~Ojo.
\newblock {Usability Evaluation of an Open Data Platform}.
\newblock In {\em Proceedings of the 18th Annual International Conference on
  Digital Government Research}, pages 495--504, 2017.

\bibitem{Otto2022}
B.~Otto, M.~tenHompel, and S.~Wrobel.
\newblock {\em {Designing Data Spaces The Ecosystem Approach to Competitive
  Advantage}}.
\newblock Springer, 2022.

\bibitem{panagiotopoulos2019public}
P.~Panagiotopoulos, B.~Klievink, and A.~Cordella.
\newblock {Public Value Creation in Digital Government}.
\newblock {\em Government Information Quarterly}, 36(4):101421, 2019.

\bibitem{pereira2020governance}
G.~V. Pereira, L.~F. Luna-Reyes, and J.~R. Gil-Garcia.
\newblock {Governance Innovations, Digital Transformation and the Generation of
  Public Value in Smart City Initiatives}.
\newblock In {\em Proceedings of the 13th International Conference on Theory
  and Practice of Electronic Governance}, pages 602--608, 2020.

\bibitem{petrova2019methodological}
D.~Petrova-Antonova and S.~Ilieva.
\newblock {Methodological Framework for Digital Transition and Performance
  Assessment of Smart Cities}.
\newblock In {\em 2019 4th International Conference on Smart and Sustainable
  Technologies (SpliTech)}, pages 1--6. IEEE, 2019.

\bibitem{pilemalm2018participatory}
S.~Pilemalm.
\newblock {Participatory Design in Emerging Civic Engagement Initiatives in the
  New Public Sector: Applying PD Concepts in Resource-Scarce Organizations}.
\newblock {\em ACM Transactions on Computer-Human Interaction}, 25(1), 2018.

\bibitem{pinheiro2020towards}
L.~Pinheiro~Junior, M.~Alexandra~Cunha, M.~Janssen, and R.~Matheus.
\newblock {Towards a Framework for Cloud Computing Use by Governments: Leaders,
  Followers and Laggers}.
\newblock In {\em The 21st Annual International Conference on Digital
  Government Research}, pages 155--163, 2020.

\bibitem{pittenger2022transformational}
L.~M. Pittenger, N.~Berente, and J.~Gaskin.
\newblock {Transformational IT Leaders and Digital Innovation: The Moderating
  Effect of Formal IT Governance}.
\newblock {\em ACM SIGMIS Database: The DATABASE for Advances in Information
  Systems}, 53(1):106--133, 2022.

\bibitem{poels2022dt4gitm}
G.~Poels, H.~A. Proper, and D.~Bork.
\newblock {DT4GITM - A Vision for a Framework for Digital Twin enabled IT
  Governance}.
\newblock In {\em Proceedings of the 55th Hawaii International Conference on
  System Sciences}, 2022.

\bibitem{potoczny2019encrypted}
I.~Potoczny-Jones, E.~Kenneally, and J.~Ruffing.
\newblock {Encrypted Dataset Collaboration: Intelligent Privacy for Smart
  Cities}.
\newblock In {\em Proceedings of the 2nd ACM/EIGSCC Symposium on Smart Cities
  and Communities}, pages 1--8, 2019.

\bibitem{IGMM}
D.~Proen{\c{c}}a, R.~Vieira, and J.~Borbinha.
\newblock {Information Governance Maturity Model Final Development Iteration}.
\newblock In {\em International Conference on Theory and Practice of Digital
  Libraries}, pages 128--139. Springer, 2017.

\bibitem{rakhman2019design}
R.~A. Rakhman, N.~Anggraini, N.~Legowo, and E.~R. Kabururuan.
\newblock {A Design of Government Enterprise Architecture Framework Based on
  G-Cloud Services}.
\newblock {\em International Journal of Scientific \& Technology Research},
  8(9):1692--1700, 2019.

\bibitem{rani2021amalgamation}
S.~Rani, R.~K. Mishra, M.~Usman, A.~Kataria, P.~Kumar, P.~Bhambri, and A.~K.
  Mishra.
\newblock {Amalgamation of Advanced Technologies for Sustainable Development of
  Smart City Environment: A Review}.
\newblock {\em IEEE Access}, 9:150060--150087, 2021.

\bibitem{reeves2022digital}
K.~Reeves, C.~Maple, and G.~Epiphaniou.
\newblock {Digital Twins - Trusted Data Environments Enabling Acceleration of
  NetZero in Critical National Infrastructure}.
\newblock In {\em Competitive Advantage in the Digital Economy}. IET, 2022.

\bibitem{reginaldo2020challenges}
F.~Reginaldo and G.~Santos.
\newblock {Challenges in Agile Transformation Journey: A Qualitative Study}.
\newblock In {\em Proceedings of the XXXIV Brazilian Symposium on Software
  Engineering}, pages 11--20, 2020.

\bibitem{reis2021exploring}
L.~C.~D. Reis, F.~C. Bernardini, S.~B. Leal~Ferreira, and C.~Cappelli.
\newblock {Exploring the Challenges of ICT Governance in Brazilian Smart
  Cities}.
\newblock In {\em Proceedings of the 14th International Conference on Theory
  and Practice of Electronic Governance}, pages 429--435, 2021.

\bibitem{ribeiro2018acceptance}
A.~Ribeiro and L.~Domingues.
\newblock {Acceptance of an Agile Methodology in the Public Sector}.
\newblock {\em Procedia Computer science}, 138:621--629, 2018.

\bibitem{rodrigues2021impacts}
G.~Rodrigues~Araujo, T.~Jose Tavares~Avila, and B.~Barreto Brasileiro~Lanza.
\newblock {Impacts of an Articulation Group for the Development of the Digital
  Government in the Brazilian Subnational Government}.
\newblock In {\em DG.O2021: The 22nd Annual International Conference on Digital
  Government Research}, pages 339--350, 2021.

\bibitem{rong2022smart}
L.~Rong.
\newblock {Smart Cities Lead the Digital Economy: Experience and Advice From
  China}.
\newblock \url{https://elib.bsu.by/bitstream/123456789/276760/1/106-117.pdf},
  2022.

\bibitem{rosler2021value}
J.~R{\"o}sler, T.~S{\"o}ll, L.~Hancock, and T.~Friedli.
\newblock {Value Co-Creation Between Public Service Organizations and the
  Private Sector: An Organizational Capabilities Perspective}.
\newblock {\em Administrative Sciences}, 11(2):55, 2021.

\bibitem{ruslan2022applying}
I.~F. Ruslan, M.~F. Alby, and M.~Lubis.
\newblock {Applying Data Governance Using DAMA-DMBOK2 Framework: The Case for
  Human Capital Management Operations}.
\newblock In {\em Proceedings of the 8th International Conference on Industrial
  and Business Engineering}, pages 336--342, 2022.

\bibitem{saba2022toward}
D.~Saba, Y.~Sahli, and A.~Hadidi.
\newblock {Toward Smart Cities Based on the Internet of Things}.
\newblock {\em Smart City Infrastructure: The Blockchain Perspective}, pages
  33--75, 2022.

\bibitem{saputro2020prerequisites}
R.~Saputro, I.~Pappel, H.~Vainsalu, S.~Lips, and D.~Draheim.
\newblock {Prerequisites for the Adoption of the X-Road Interoperability and
  Data Exchange Framework: A Comparative Study}.
\newblock In {\em 2020 7th International Conference on eDemocracy \&
  eGovernment (ICEDEG)}, pages 216--222. IEEE, 2020.

\bibitem{sarah2020devops}
A.~Sarah and B.~Fakieh.
\newblock {How DevOps Practices Support Digital Transformation}.
\newblock {\em International Journal}, 9(3), 2020.

\bibitem{saura2022handbook}
J.~R. Saura and F.~Debasa.
\newblock {\em {Handbook of Research on Artificial Intelligence in Government
  Practices and Processes}}.
\newblock IGI Global, 2022.

\bibitem{schedler2019smart}
K.~Schedler, A.~A. Guenduez, and R.~Frischknecht.
\newblock {How Smart Can Government Be? Exploring Barriers to the Adoption of
  Smart Government}.
\newblock {\em Information Polity}, 24(1):3--20, 2019.

\bibitem{scrupola2021value}
A.~Scupola and I.~Mergel.
\newblock {Value Co-Creation and Digital Service Transformation: The Case of
  Denmark}.
\newblock {\em Available at SSRN 3896269}, 2021.

\bibitem{shore2008art}
J.~Shore and S.~Warden.
\newblock {\em {The Art of Agile Development}}.
\newblock Theory in practice. O'Reilly Media, Incorporated, 2008.

\bibitem{Shiny_apps}
{Shyny from RStudio}.
\newblock {Shiny: R package That Makes It Easy to Build Interactive Web Aps
  Straight From R}.
\newblock \url{https://shiny.rstudio.com/}.
\newblock [Online; accessed 17-September-2022].

\bibitem{soe2017finest}
R.-M. Soe.
\newblock {FINEST Twins: Platform For Cross-Border Smart City Solutions}.
\newblock In {\em Proceedings of the 18th Annual International Conference on
  Digital Government Research}, pages 352--357, 2017.

\bibitem{soe2017smart}
R.-M. Soe.
\newblock {Smart Twin Cities Via Urban Operating System}.
\newblock In {\em Proceedings of the 10th International Conference on Theory
  and Practice of Electronic Governance}, pages 391--400, 2017.

\bibitem{sousa2018liberopinion}
A.~A. Sousa, P.~Agante, S.~Abrantes, and L.~B. Gouveia.
\newblock {Liberopinion: A Web Platform for Public Participation}.
\newblock In {\em Proceedings of the 11th International Conference on Theory
  and Practice of Electronic Governance}, ICEGOV '18, page 199–208, New York,
  NY, USA, 2018. ACM.

\bibitem{Nat_CS_Agency_ES}
{Spain Government}.
\newblock {Spanish National Cybersecurity Institute (INCIBE)}.
\newblock \url{https://www.incibe.es/en/}.
\newblock [Online; accessed 05-September-2022].

\bibitem{stirbu2018towards}
V.~Stirbu and T.~Mikkonen.
\newblock {Towards Agile Yet Regulatory-Compliant Development of Medical
  Software}.
\newblock In {\em 2018 IEEE International Symposium on Software Reliability
  Engineering Workshops (ISSREW)}, pages 337--340. IEEE, 2018.

\bibitem{syahrizal2020finding}
A.~Syahrizal, A.~Arief, and D.~I. Sensuse.
\newblock {Finding Preeminence: A Systematic Literature Review of E-Service
  Success Factors}.
\newblock In {\em 2020 5th International Conference on Informatics and
  Computing (ICIC)}, pages 1--6. IEEE, 2020.

\bibitem{tambouris2021towards}
E.~Tambouris and K.~Tarabanis.
\newblock {Towards Inclusive Integrated Public Service (IPS) Co-Creation and
  Provision}.
\newblock In {\em DG.O2021: The 22nd Annual International Conference on Digital
  Government Research}, pages 458--462, 2021.

\bibitem{tamburri2020dataops}
D.~A. Tamburri, W.-J. Van Den~Heuvel, and M.~Garriga.
\newblock {Dataops for Societal Intelligence: A Data Pipeline for Labor Market
  Skills Extraction and Matching}.
\newblock In {\em 2020 IEEE 21st International Conference on Information Reuse
  and Integration for Data Science (IRI)}, pages 391--394. IEEE, 2020.

\bibitem{tangi2020barriers}
L.~Tangi, M.~Janssen, M.~Benedetti, and G.~Noci.
\newblock {Barriers and Drivers of Digital Transformation in Public
  Organizations: Results From a Survey in the Netherlands}.
\newblock In {\em International Conference on Electronic Government}, pages
  42--56. Springer, 2020.

\bibitem{tangi2021digital}
L.~Tangi, M.~Janssen, M.~Benedetti, and G.~Noci.
\newblock {Digital Government Transformation: A Structural Equation Modelling
  Analysis of Driving and Impeding Factors}.
\newblock {\em International Journal of Information Management}, 60:102356,
  2021.

\bibitem{teixeira2020maturity}
D.~Teixeira, R.~Pereira, T.~Henriques, M.~M.~D. Silva, J.~Faustino, and
  M.~Silva.
\newblock {A Maturity Model for DevOps}.
\newblock {\em International Journal of Agile Systems and Management},
  13(4):464--511, 2020.

\bibitem{Data_Authority_IT}
{The Italian Data Protection Authority}.
\newblock {The Italian Data Protection Authority Website}.
\newblock \url{https://www.garanteprivacy.it/web/garante-privacy-en/home_en}.
\newblock [Online; accessed 10-October-2022].

\bibitem{New_Yorker}
{The New Yorker Magazine}.
\newblock {Estonia, the Digital Republic}.
\newblock
  \url{https://www.newyorker.com/magazine/2017/12/18/estonia-the-digital-republic}.
\newblock [Online; accessed 10-January-2023].

\bibitem{GovTech}
{The World Bank}.
\newblock {(GovTech: The New Frontier in Digital Government Transformation)}.
\newblock \url{https://www.worldbank.org/en/programs/govtech}.
\newblock [Online; accessed 30-September-2022].

\bibitem{Dig_Adop_Index}
{The World Bank}.
\newblock {World Bank Digital Adoption Index}.
\newblock
  \url{https://www.worldbank.org/en/publication/wdr2016/Digital-Adoption-Index}.
\newblock [Online; accessed 30-September-2022].

\bibitem{w3consortium}
J.~J. Tim Berners-Lee.
\newblock {World Wide Web Consortium}.
\newblock \url{https://w3c.org/}, 2022.
\newblock [Online; accessed 03-June-2022].

\bibitem{Virtual_Singapore}
{Tomorrow City: The Heart of Urban Innovation}.
\newblock {Virtual Singapore: Singapore Experiments With Its Digital Twin to
  Improve City Life}.
\newblock
  \url{https://www.tomorrow.city/a/singapore-experiments-with-its-digital-twin-to-improve-city-life}.
\newblock [Online; accessed 15-December-2022].

\bibitem{toots2017open}
M.~Toots, K.~Mcbride, T.~Kalvet, and R.~Krimmer.
\newblock {Open Data As Enabler of Public Service Co-Creation: Exploring the
  Drivers and Barriers}.
\newblock In {\em 2017 Conference for E-Democracy and Open Government (CeDEM)},
  pages 102--112. IEEE, 2017.

\bibitem{toots2017framework}
M.~Toots, K.~McBride, T.~Kalvet, R.~Krimmer, E.~Tambouris, E.~Panopoulou,
  E.~Kalampokis, and K.~Tarabanis.
\newblock {A Framework for Data-Driven Public Service Co-Production}.
\newblock In {\em International Conference on Electronic Government}, pages
  264--275. Springer, 2017.

\bibitem{umar2022digital}
A.~Umar.
\newblock {A Digital Transformation Lab for Developing Countries and Small to
  Medium Enterprises}.
\newblock In {\em 2022 IEEE European Technology and Engineering Management
  Summit (E-TEMS)}, pages 160--165. IEEE, 2022.

\bibitem{UN_Index}
{United Nations}.
\newblock {United Nations E-Government Development Index}.
\newblock
  \url{https://publicadministration.un.org/egovkb/en-us/About/Overview/-E-Government-Development-Index}.
\newblock [Online; accessed 30-September-2022].

\bibitem{Nat_CS_Agency_US}
{United States Government}.
\newblock {Cybersecurity \& Infrastructure Agency (CISA)}.
\newblock \url{https://www.cisa.gov/}.
\newblock [Online; accessed 05-September-2022].

\bibitem{urbanczyk2021application}
W.~Urbanczyk and J.~Werewka.
\newblock {Application of a Government Data Center (GDC) Reference Model for
  Security Management Analysis}.
\newblock In {\em 2021 IEEE International Conference on E-Business Engineering
  (ICEBE)}, pages 112--119. IEEE, 2021.

\bibitem{AdvDigEquityAll2022}
{US Department of Education - Office of Educational Technology}.
\newblock {Advancing Digital Equity for All: Community-Based Recommendations
  for Developing Effective Digital Equity Plans to Close the Digital Divide and
  Enable Technology-Empowered Learning}.
\newblock
  \url{https://tech.ed.gov/files/2022/09/DEER-Resource-Guide_FINAL.pdf}.
\newblock [Online; accessed 20-March-2023].

\bibitem{vaira2022smart}
V.~Vaira.
\newblock {Smart City Governance and the Challenge of Digital Platforms Within
  the Public Sector}.
\newblock In {\em 2022 IEEE International Smart Cities Conference (ISC2)},
  pages 1--7, 2022.

\bibitem{E-Gov-MM}
G.~Valdés, M.~Solar, H.~Astudillo, M.~Iribarren, G.~Concha, and M.~Visconti.
\newblock {Conception, Development and Implementation of an E-Government
  Maturity Model in Public Agencies}.
\newblock {\em Government Information Quarterly}, 28(2):176--187, 2011.

\bibitem{van2020evaluating}
C.~van Noordt and G.~Misuraca.
\newblock {Evaluating the Impact of Artificial Intelligence Technologies in
  Public Services: Towards an Assessment Framework}.
\newblock In {\em Proceedings of the 13th International Conference on Theory
  and Practice of Electronic Governance}, pages 8--16, 2020.

\bibitem{CKanDMS}
{Various Authors}.
\newblock {Comprehensive Knowledge Archive Network (CKan): The Open-Source Data
  Management System for Powering Data Hubs and Data Portals.}
\newblock \url{https://ckan.org/}.
\newblock [Online; accessed 18-August-2022].

\bibitem{DECODEproject}
{Various Authors}.
\newblock {European DECODE Project}.
\newblock \url{https://decodeproject.eu/}.
\newblock [Online; accessed 18-August-2022].

\bibitem{SENTILOproject}
{Various Authors}.
\newblock {Sentilo: An Open-Source Software Sensor and Actuator Project}.
\newblock \url{https://sentilo.io/}.
\newblock [Online; accessed 18-August-2022].

\bibitem{vestues2021usingPeople}
K.~Vestues.
\newblock {\em {Using Digital Platforms to Promote Value Co-Creation: A Case
  Study of a Public Sector Organization}}.
\newblock PhD thesis, NTNU, 2021.

\bibitem{vestues2021usingProcess}
K.~Vestues, M.~Mikalsen, and E.~Monteiro.
\newblock {Using Digital Platforms to Promote a Service-Oriented Logic in
  Public Sector Organizations: A Case Study}.
\newblock In {\em Proceedings of the 54th Hawaii International Conference on
  System Sciences}, page 2193, 2021.

\bibitem{vial2019understanding}
G.~Vial.
\newblock {Understanding digital transformation: A review and a research
  agenda}.
\newblock {\em The Journal of Strategic Information Systems}, 28(2):118--144,
  2019.

\bibitem{vijai2020cloud}
C.~Vijai.
\newblock {Cloud-Based E-Governance in India}.
\newblock {\em International Journal of Management}, 2020.

\bibitem{viscusi2020governments}
G.~Viscusi, A.~Collins, and M.-V. Florin.
\newblock {Governments' Strategic Stance Toward Artificial Intelligence: An
  Interpretive Display on Europe}.
\newblock In {\em Proceedings of the 13th International Conference on Theory
  and Practice of Electronic Governance}, pages 44--53, 2020.

\bibitem{vogelsang2019taxonomy}
K.~Vogelsang, K.~Liere-Netheler, S.~Packmohr, and U.~Hoppe.
\newblock {A Taxonomy of Barriers to Digital Transformation}.
\newblock In {\em 14th International Conference on Wirtschaftsinformatik,
  Siegen, Germany (February 24-27, 2019)}, pages 736--750. Universit{\"a}t
  Siegen, 2019.

\bibitem{volpe2022supporting}
M.~Volpe, I.~G. Rojas, G.~Gaffuri, R.~Marfievici, E.~Genova, A.~Gheorghe,
  J.~Kniewallner, and O.~Veledar.
\newblock {Supporting Innovation in Smart Cities Through Cascade Funding: The
  Case of Water Management}.
\newblock In {\em 2022 IEEE International Smart Cities Conference (ISC2)},
  pages 1--7, 2022.

\bibitem{wang2023digital}
H.~Wang, X.~Chen, F.~Jia, and X.~Cheng.
\newblock {Digital Twin-Supported Smart City: Status, Challenges and Future
  Research Directions}.
\newblock {\em Expert Systems With Applications}, page 119531, 2023.

\bibitem{wingren2021blockchain}
J.~Wingren and Z.~Wes{\'e}n.
\newblock {Blockchain in the Public Sector: Applications For Improving Services
  in Society}, 2021.

\bibitem{wipulanusat2019drivers}
W.~Wipulanusat, K.~Panuwatwanich, R.~A. Stewart, and J.~Sunkpho.
\newblock {Drivers and Barriers to Innovation in the Australian Public Service:
  A Qualitative Thematic Analysis}.
\newblock {\em Engineering Management in Production and Services}, 11(1):7--22,
  2019.

\bibitem{wolff2019will}
C.~Wolff, A.~Omar, and Y.~Shildibekov.
\newblock {How Will We Build Competences For Managing the Digital
  Transformation?}
\newblock In {\em 2019 10th IEEE International Conference on Intelligent Data
  Acquisition and Advanced Computing Systems: Technology and Applications
  (IDAACS)}, volume~2, pages 1122--1129. IEEE, 2019.

\bibitem{Semantic_Web}
{World Wide Web Consortium (W3C)}.
\newblock {Semantic Web}.
\newblock \url{https://www.w3.org/standards/semanticweb/}.
\newblock [Online; accessed 03-October-2022].

\bibitem{yarlagadda2018public}
R.~T. Yarlagadda.
\newblock {How Public Sectors Can Adopt the DevOps Practices to Enhance the
  System}.
\newblock {\em International Journal of Emerging Technologies and Innovative
  Research (www. jetir. org - UGC and ISSN Approved), ISSN}, pages 2349--5162,
  2018.

\bibitem{ylinen2021digital}
M.~Ylinen.
\newblock {\em {Digital Transformation in a Finnish Municipality: Tensions as
  Drivers of Continuous Change}}.
\newblock PhD thesis, Tampere University, 2021.

\bibitem{ylinen2021incorporating}
M.~Ylinen.
\newblock {Incorporating Agile Practices in Public Sector IT Management: A
  Nudge Toward Adaptive Governance}.
\newblock {\em Information Polity}, 26(3):251--271, 2021.

\bibitem{ylipulli2020smart}
J.~Ylipulli and A.~Luusua.
\newblock {Smart Cities With a Nordic Twist? Public Sector Digitalization in
  Finnish Data-Rich Cities}.
\newblock {\em Telematics and Informatics}, 55:101457, 2020.

\bibitem{young2017civic}
M.~Young and A.~Yan.
\newblock {Civic Hackers’ User Experiences and Expectations of Seattle’s
  Open Municipal Data Program}.
\newblock In {\em Proceedings of the 50th Hawaii International Conference on
  System Sciences}, 2017.

\bibitem{yukhno2022digital}
A.~Yukhno.
\newblock {Digital Transformation: Exploring Big Data Governance in Public
  Administration}.
\newblock {\em Public Organization Review}, pages 1--15, 2022.

\bibitem{zazour2020devops}
M.~Zarour, N.~Alhammad, M.~Alenezi, and K.~Alsarayrah.
\newblock {Devops Process Model Adoption in Saudi Arabia: An Empirical Study}.
\newblock {\em Jordanian Journal of Computers and Information Technology
  (JJCIT)}, 6(03), 2020.

\bibitem{zeleti2019agile}
F.~A. Zeleti and A.~Ojo.
\newblock {Agile Mechanisms For Open Data Process Innovation in Public Sector
  Organizations: Towards Theory Building}.
\newblock In {\em Proceedings of the 12th International Conference on Theory
  and Practice of Electronic Governance}, pages 164--174, 2019.

\bibitem{zhang2022framework}
J.~Zhang, C.~Chen, Y.~Zhang, Y.~Cui, P.~Han, N.~Meng, and Y.~Xu.
\newblock {The Framework and Practices of Digital Twin City}.
\newblock In {\em 2022 IEEE 12th International Conference on Electronics
  Information and Emergency Communication (ICEIEC)}, pages 111--116. IEEE,
  2022.

\bibitem{zheng2020towards}
Y.~Zheng, A.~Pal, S.~Abuadbba, S.~R. Pokhrel, S.~Nepal, and H.~Janicke.
\newblock {Towards IoT Security Automation and Orchestration}.
\newblock In {\em the 2nd IEEE International Conference on Trust, Privacy and
  Security in Intelligent Systems and Applications (TPS-ISA)}, pages 55--63.
  IEEE, 2020.

\end{thebibliography}

\end{document}